\documentclass[journal,final,twocolumn,10pt,twoside]{IEEEtran}

\normalsize

\IEEEoverridecommandlockouts 

\ifCLASSINFOpdf

\else
\fi

\usepackage{mathtools}
\usepackage{amsmath}
\usepackage{amssymb}
\usepackage{amsmath}
\usepackage{cite}
\usepackage{siunitx}
\usepackage{multirow}
\usepackage{multicol}
\DeclareMathOperator{\E}{\mathbb{E}}
\usepackage{graphicx}
\usepackage{color}
\usepackage{xcolor}
\date{}

\usepackage{calc}
\newcolumntype{M}[1]{>{\centering\arraybackslash}m{#1}}
\newcolumntype{N}{@{}m{0pt}@{}}

\usepackage{mdwmath}
\usepackage{blindtext}
\usepackage{eqparbox}
\usepackage{fixltx2e}
\begin{document}
\title{Uplink Non-Orthogonal Multiple Access over Mixed RF-FSO Systems}
\author{Mohammad Vahid Jamali, and Hessam Mahdavifar,~\IEEEmembership{\normalsize Member,~IEEE}
	\thanks{The authors are with the Electrical Engineering and Computer Science Department, University of Michigan, Ann Arbor, MI, USA (e-mail: mvjamali@umich.edu and hessam@umich.edu).}
\thanks{This work was presented in part at the IEEE Global Communications Conference (GLOBECOM), Abu Dhabi,
UAE, Dec. 2018 \cite{jamali2018outage}. This work was supported by the National Science Foundation under grants CCF--1763348, and CCF--1909771.}
}
\maketitle
\begin{abstract}
\boldmath 
In this paper, we consider a relay-assisted uplink non-orthogonal multiple access (NOMA) system. In this system, two radio frequency (RF) users are grouped for simultaneous transmissions, over each resource block, to an intermediate relay. The relay then forwards the amplified version of the users' aggregated signals, in the presence of multiuser interference, to a relatively far destination. In order to cope with the users' ever-increasing desire for higher data rates, a high-throughput free-space optics (FSO) link is employed as the relay-destination backhaul link. It is assumed that the FSO backhaul link is subject to Gamma-Gamma turbulence with pointing error. Also, a Rayleigh fading model is considered for the user-relay access links. Under these assumptions, we derive closed-form expressions for the outage probability and tractable forms, involving only one-dimensional integrals, for the ergodic capacity.
 Moreover, the outage probability and ergodic capacity analysis are extended to the conventional RF-backhauled systems in the presence of multiuser interference to both relay and destination nodes, and Rician fading for the relay-destination RF link.
Our results reveal the superiority of FSO backhauling for high-throughput and high-reliability NOMA systems compared to RF backhauling. This work can be considered as a general analysis of dual-hop uplink NOMA systems as well as the first attempt to incorporate power-domain NOMA in mixed RF-FSO systems.
\end{abstract}
\section{Introduction}
\IEEEPARstart{N}{on-orthogonal} multiple access (NOMA) is widely considered as one of the enabling technologies for the fifth generation (5G) wireless networks. With its two general  power- and code-domain forms, NOMA can potentially pave the way toward higher throughput, lower latency, improved fairness, higher reliability, and massive connectivity \cite{dai2015non}. Motivated by these fascinating advantages, extensive research activities have been carried out in the past few years to advance NOMA in diverse directions \cite{ding2017survey,islam2017power}. 

Excavating the rich literature on NOMA, some research activities have focused on the code-domain NOMA which, generally speaking, attempts to serve a set of users in a smaller set of orthogonal resource blocks using a pattern matrix. In this context, variety of multiple access mechanisms have been proposed such as sparse code multiple access (SCMA) \cite{nikopour2013sparse}, lattice partition multiple access (LPMA) \cite{qiu2018lattice}, interleave-grid multiple access (IGMA) \cite{xiong2017advanced}, and pattern division multiple access (PDMA) \cite{chen2017pattern}. However, code-domain NOMA usually suffers from a high detection complexity due to the need for complex multiuser detection methods such as maximum likelihood (ML) detection, massage passing algorithm (MPA), and maximum  \textit{a posteriori} (MAP) detection.
To this end, some of the recent work in code-domain NOMA have focused on lowering the detection complexity, e.g., using list sphere
decoding \cite{wei2016low}, and recursive detection approaches enabled by  sophisticated designs of the overall pattern matrix \cite{jamali2018low}.

Power-domain NOMA, on the other hand, has attracted more attention from the research and industrial communities because of its relative simplicity. For example, a two-user downlink power-domain NOMA system, also referred to as multiuser superposition transmission, has been proposed for inclusion in the
Third Generation Partnership Project (3GPP) long-term evolution
advanced (LTE-A) standard \cite{must}. Motivated by this, power-domain NOMA has been explored in various directions, including multiple-input multiple-output systems \cite{sun2015ergodic}, cooperative transmission \cite{ding2015cooperative}, simultaneous wireless information and power transfer \cite{liu2016cooperative,chang2018energy,do2017bnbf}, ultra-reliable and low-latency communications \cite{amjad2018performance}, visible-light communications \cite{kizilirmak2015non}, and millimeter-wave (mmWave) communications \cite{ding2017random}, (see, e.g., \cite{ding2017survey} for a comprehensive survey). 

In a variety of applications, there is a need to transmit the users' data to a central unit or a wired base station (BS); however, given the limited power of the users, it is not feasible for the users to directly communicate with the relatively far destination. To this end, several recent works have considered the relaying problem in downlink and uplink NOMA communications. In particular, capacity analysis of a simple cooperative relaying system, consisting of a source, a relay, and a destination node is provided in \cite{kim2015capacity}. The outage probabilities and ergodic sum rate of a downlink two-user NOMA system, with a full-duplex relay helping one of the users, are characterized in \cite{zhong2016non}. Performance of downlink NOMA transmission with an intermediate amplify-and-forward (AF) relay for multiple-antenna systems, and over Nakagami-$m$ fading channels is investigated in \cite{men2015non} and \cite{men2017performance}, respectively. The performance of coordinated direct and relay transmission for two-user downlink and uplink NOMA systems is investigated in \cite{kim2015coordinated} and \cite{kader2018coordinated}, respectively. Hybrid decode-and-forward (DF) and AF relaying in NOMA systems is proposed in \cite{liu2016hybrid}, and forwarding strategy selection (either AF or DF) problem is explored in \cite{xiao2018forwarding}. Moreover, a comprehensive performance evaluation of a two-user downlink cooperative NOMA system is provided in \cite{yue2018exploiting}, where one of the users acts as a relay switching between half-duplex and full-duplex modes. In addition, a dynamic DF-based cooperative scheme has been proposed in \cite{zhou2018dynamic} for downlink NOMA transmission with spatially random users.

The aforementioned prior works often assume communications in the absence of external multiuser interference to the NOMA users and the conventional sub-6 GHz radio frequency (RF) for the backhaul links.
  However, the available bandwidth in the sub-6 GHz band is scarce and falls short of supporting the users' aggressive demand for the higher data rates, especially when NOMA is employed in the user-relay access links to provide higher throughputs. In this case, the relay-destination backhaul link can pose a severe bottleneck on the end-to-end performance and substantially negate the NOMA advantages through reducing  the users' achievable throughput and reliability which can in turn even increase their latency.

A potential approach to overcome the aforementioned drawback is to utilize higher frequency bands, e.g., through the deployment of mmWave and/or free-space optics (FSO) backhaul links \cite{ge20145g,demers2011survey}. MmWave communication is usually preferred for relatively short communication lengths due to the severe propagation conditions at millimeter frequencies \cite{andrews2017modeling}. FSO links, on the other hand, can provide much more available bandwidth and support ranges in the order of several kilometers \cite{khalighi2014survey}. 
To this end, in this paper, we investigate the performance of uplink NOMA transmission over mixed RF-FSO systems. In particular, an AF relay is employed to forward the amplified received signal from the Rayleigh fading access links to the destination through an ultra high-throughput directive interference-free FSO link subject to Gamma-Gamma (GG) fading with beam misalignment error. This paper can be considered as a general analysis of dual-hop uplink NOMA systems, and also an initial attempt to incorporate power-domain NOMA in mixed RF-FSO systems. 

Our main contributions can be summarized as follows. 
\begin{itemize}
	\item We consider a general dual-hop uplink NOMA transmission subject to the presence of multiuser interference from some independent users. Such interference can be induced, e.g., due the co-channel interference from  nearby users aiming to communicate to other relays or destinations. The inclusion of external interference can be also helpful in the analysis of mmWave NOMA, where side-lobes of nearby mmWave beams cause inter-beam interference to the power-domain NOMA users grouped over a given mmWave beam \cite{cui2018optimal,wei2019multi}. 
\item We apply dynamic-order decoding
to determine the detection order of the NOMA users at the destination. 
\item We derive the closed-form expressions for the individual- and sum-rate outage probabilities of the mixed RF-FSO uplink NOMA system with respect to dynamic-order decoding at the destination, AF scheme at the relay, Rayleigh fading for the user-relay access links, and GG turbulence with the inclusion of pointing error for the relay-destination backhaul FSO link. 
\item We further derive the outage probability closed forms (in terms of an infinite series that can effectively be approximated by some finite number of terms) for the RF-backhauled system when the relay-destination backhaul link is subject to Rician fading, and both relay and destination nodes are subject to external multiuser interference.
\item Average individual- and sum-rate formulas are characterized for both FSO- and RF-backhauled systems up to only one-dimensional integrals over the fading coefficient of the backhaul link. That is equivalent to say that the ergodic capacity closed-form expressions are obtained for the single-hop uplink NOMA subject to some exterior multiuser interference, or better to say, for the aforementioned dual-hop system model given each realization of the backhaul fading coefficient. 
\item Extensive numerical results are provided to validate the accuracy of the derived formulas and also ascertain the system performance  over different channel conditions.
\end{itemize}

The rest of the paper is organized as follows. In Section II, we describe the system model. In Section III, we derive the individual- and sum-rate outage probability closed-form formulas for both FSO- and RF-backhauled dual-hop uplink NOMA systems. Section IV is devoted to the ergodic capacity analysis for the same system model, Section V provides the numerical results, and Section VI concludes the paper. 
\begin{figure}\label{BDfig}
	\centering
\includegraphics[width=3.4in, trim = {0 0 0 0}, clip]{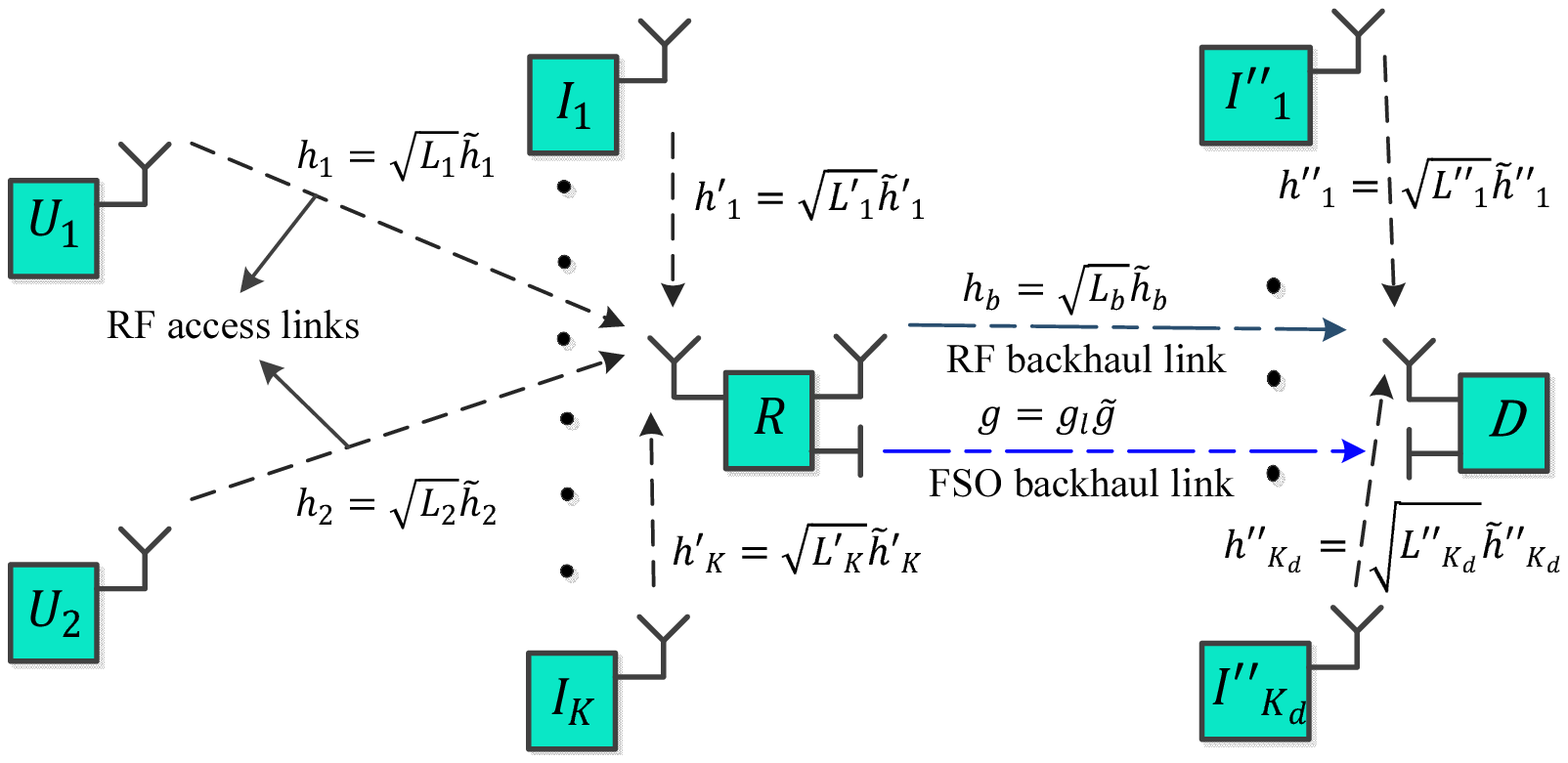}
	\caption{Block diagram of the uplink NOMA transmission over mixed RF-FSO and dual-hop RF/RF systems.}
	\vspace{-0.15in}
\end{figure}
\section{System Model}
As shown in Fig. 1, consider two RF users $\mathcal{U}_1$ and  $\mathcal{U}_2$ grouped together for uplink NOMA transmission to an AF relay $\mathcal{R}$. Denote the composite RF channel gain of the $\mathcal{U}_i-\mathcal{R}$ link by $h_i=\sqrt{L_i}\tilde{h}_i$, $i=1,2$, where $L_i$ and $\tilde{h}_i$ are the path-loss gain and the independent-and-identically-distributed (i.i.d.) Rayleigh fading coefficient of the $\mathcal{U}_i-\mathcal{R}$ RF link, respectively. 
{The path-loss gain is given by $L_i = {G^{\mathrm{RF}}_{t,i} G^{\mathrm{RF}}_{r,\mathcal{R}}}\times\left[{\lambda^{\mathrm{RF}}}/{(4\pi d^{\mathrm{RF}}_{\mathrm{ref}})}\right]^2 \times \left[{d^{\mathrm{RF}}_{\mathrm{ref}}}/{d_i^{\mathrm{RF}}}\right]^\nu$ \cite[Eq. (2)]{jamali2016link}
	in which $\lambda^{\mathrm{RF}}$ is the wavelength of the RF signal, $G^{\mathrm{RF}}_{t,i}$ and $G^{\mathrm{RF}}_{r,\mathcal{R}}$ are the RF transmit and receive
	antenna gains of the $\mathcal{U}_i-\mathcal{R}$ link, respectively, $d^{\mathrm{RF}}_{\mathrm{ref}}$ is a reference distance for the antenna far-field, $d^{\mathrm{RF}}_i$ is the $\mathcal{U}_i-\mathcal{R}$ link length, and $\nu$ is the RF path-loss exponent.}
 Furthermore, assume that the uplink transmission to the relay is affected by undesired multiuser interference from $K$ interfering users $\mathcal{I}_k$, $k=1,2,...,K$, each with the transmit power $p'_k$, path-loss gain $L'_k$, and i.i.d. Rayleigh fading coefficient $\tilde{h}'_k$. As explained in Section I, this interference can be from the users scheduled for the concurrent transmission to some other relays in the cellular network or any other non-vanishing interference during the desired transmission block. 
The  received signal  by the relay can then be expressed as
\begin{align}\label{y_R}
y_{\mathcal{R}}=\sum_{i=1}^{2}x_i\tilde{h}_i\sqrt{a_iL_iP}+\sum_{k=1}^{K}x'_k\tilde{h}'_k\sqrt{L'_kp'_k}+n_{\mathcal{R}},
\end{align}
where $x_i$ and $x'_k$ are the transmit symbols by $\mathcal{U}_i$ and $\mathcal{I}_k$, respectively,
$a_1$ and $a_2=1-a_1$ are the power coefficients, and $n_{\mathcal{R}}$ is the additive white Gaussian noise (AWGN) of the relay receiver with mean zero and variance $\sigma^2_{\mathcal{R}}$. 
 Note that for the users with independent Rayleigh fading, all fading gains $|\tilde{h}_i|^2$'s and $|\tilde{h}'_k|^2$'s have an exponential distribution with mean one (to ensure that fading neither amplifies nor attenuates the received power) as $f_{|\tilde{h}_i|^2}(x)=f_{|\tilde{h}'_k|^2}(x)=\exp(-x)$, $x\geq 0$.

The received signal $y_{\mathcal{R}}$ at the relay is then converted to optical signal using intensity-modulation direct-detection (IM/DD), and is amplified with a constant gain $G$ to keep the disparity between the power levels of different NOMA users for successive interference cancellation (SIC) detection at the destination. In this case, the transmitted optical signal by the relay toward the destination $\mathcal{D}$ can be expressed as $S_{\mathcal{R}}=G(1+\eta y_{\mathcal{R}})$, where $\eta$ is the electrical-to-optical conversion coefficient \cite{lee2011performance}.
The transmitted signal then undergoes the  FSO channel with the composite gain $g=g_l\tilde{g}$ where $g_l$ is the path-loss gain of the $\mathcal{R-D}$ FSO backhaul link, with the length $d_{\mathcal{RD}}$, defined as $g_l=\rho\times10^{-\kappa d_{\mathcal{RD}}/10}$
where $\rho$ is the responsivity of the photodetector, and $\kappa$ is the weather-dependent attenuation coefficient \cite{jamali2016link}.
 Moreover, $\tilde{g}=g_pg_f$ is the total fading coefficient due to pointing error $g_p$ and optical turbulence $g_f$. In the case of GG optical turbulence with beam misalignment, the distribution of $\tilde{g}$ can be expressed as \cite{sandalidis2009optical}
 \begin{align}\label{f_gtild}
\! f_{\tilde{g}}(\tilde{g})=\frac{\alpha\beta\xi^2}{A_0\Gamma(\alpha)\Gamma(\beta)}{\rm G}_{1,3}^{3,0}\left[\frac{\alpha\beta}{A_0}\tilde{g}{\bigg |\begin{matrix}
 \xi^2	\\ \xi^2-1,\alpha\!-\!1,\beta\!-\!1
 	\end{matrix}}\right],
 \end{align}
 where $\alpha$ and $\beta$ are the fading parameters of the GG distribution, $\xi$ is the ratio of the equivalent beam radius and the pointing error
displacement standard deviation (jitter) measured at the receiver, $\Gamma(\cdot)$ is the gamma function \cite[Eq. (8.310)]{gradshteyn2014table}, and ${\rm G}[\cdot]$ is the Meijer's G-function \cite[Eq. (9.301)]{gradshteyn2014table}. Furthermore, $A_0$ is the geometric loss in the case of perfect beam alignment (zero radial displacement) defined as $A_0=[{\rm erf}(\sqrt{\pi}r/(\sqrt{2}\phi d_{\mathcal{RD}}))]^2$ in which ${\rm erf}(\cdot)$ is the error function, $r$ is the receiver aperture radius, and $\phi$ is the transmitter beam divergence angle. 

The destination then filters out the direct current (DC) component of $g_l\tilde{g}G$ from $g_l\tilde{g}S_{\mathcal{R}}+n_{\mathcal{D}}$ to obtain the received signal as
\begin{align}\label{y_D}
y_{\mathcal{D}}=&\eta g_l\tilde{g}G\bigg(\sum_{i=1}^{2}x_i\tilde{h}_i\sqrt{a_iL_iP}
+\sum_{k=1}^{K}x'_k\tilde{h}'_k\sqrt{L'_kp'_k}+n_{\mathcal{R}}\bigg)\nonumber\\&+n_{\mathcal{D}},
\end{align}
 where $n_{\mathcal{D}}$ is the destination AWGN with mean zero and variance $\sigma^2_{\mathcal{D}}$.
 
 We assume that the NOMA users are indexed based on their path-loss gains, i.e., $L_1\geq L_2$, and the power allocation strategy proposed in \cite{zhang2016uplink} is adopted to determine $a_1$ and $a_2$ as $a_1L_1=a_2L_2\times10^{s/10}$ where $s\geq 0$ is the power back-off step; hence, $a_1=L_2\times10^{s/10}/(L_1+L_2\times10^{s/10})$ and $a_2=L_1/(L_1+L_2\times10^{s/10})$. 
{We further consider dynamic-order decoding at the destination \cite{gao2017theoretical,najafi2018non}, that is expected to achieve a higher performance compared to fixed-order decoding assuming that the BS has perfect knowledge about the channel state information (CSI) and sorts the NOMA users based on their instantaneous received power.}
 In fact, based on the principles of uplink power-domain NOMA \cite{zhang2016uplink,yang2016general}, the BS sorts the users based on their channel conditions from the best to the worst.
 Therefore, depending on the fading coefficients $\tilde{h}_1$ and $\tilde{h}_2$, the detection order is $\pi_1=(1,2)$, meaning that the first user is decoded first, if $a_1L_1|\tilde{h}_1|^2\geq a_2L_2|\tilde{h}_2|^2$; otherwise, the detection order is $\pi_2=(2,1)$. 
 \section{Outage Probability Analysis}
In this section, we first characterize the individual- and sum-rate outage probabilities for mixed RF-FSO NOMA systems and then extend the results to the RF-backhauled systems. 
\subsection{Individual-Rate Outage Analysis}
Note that if the detection order is $\pi_1$, the SIC receiver first treats the signal from the second NOMA user as noise to decode $x_1$ with the signal-to-interference-plus-noise ratio (SINR) given by
\begin{align}\label{gamma_f}
\!\!\gamma^{(1)}_{\pi_1}\!=\!\frac{a_1L_1P\tilde{g}^2|\tilde{h}_1|^2}{a_2L_2P\tilde{g}^2|\tilde{h}_2|^2\!+\!\sum_{k=1}^{K}{L'_kp'_k}\tilde{g}^2|\tilde{h}'_k|^2\!+\!\tilde{g}^2\sigma^2_{\mathcal{R}}\!+\!C_{\mathcal{D}}},
\end{align}
and then, after removing the received power from the first user, decodes $x_2$ with the SINR given by
\begin{align}\label{gamma_s}
\gamma^{(2)}_{\pi_1}=\frac{a_2L_2P\tilde{g}^2|\tilde{h}_2|^2}{\sum_{k=1}^{K}{L'_kp'_k}\tilde{g}^2|\tilde{h}'_k|^2+\tilde{g}^2\sigma^2_{\mathcal{R}}+C_{\mathcal{D}}},
\end{align}
where $C_{\mathcal{D}}\triangleq\sigma^2_{\mathcal{D}}/(\eta^2 g_l^2G^2)$. Similarly, when the detection order is $\pi_2$  the SINR values $\gamma^{(1)}_{\pi_2}$ and $\gamma^{(2)}_{\pi_2}$ can be obtained by properly changing the indices in \eqref{gamma_f} and \eqref{gamma_s}.

\begin{figure*}[!t]
	\normalsize
	\vspace{-0.05cm}
	\setcounter{equation}{8}
	\begin{align}\label{Joint1,1}
	\Pr(\gamma^{(1)}_{\pi_1}<\gamma^{(1)}_{\rm th},\pi_1){\Big|}_{\gamma^{(1)}_{\rm th}<1}
	&	\stackrel{(a)}{=}\Pr\left(|\tilde{h}_1|^2<\gamma^{(1)}_{\rm th}\left[|\tilde{h}_2|^2\times10^{-s/10}+{\boldsymbol{\mathcal{I}}_1}+C_{\mathcal{D}}/(a_1L_1P\tilde{g}^2)\right],|\tilde{h}_1|^2\geq |\tilde{h}_2|^2\times10^{-s/10}\right)\nonumber\\
	&\hspace{-3.8cm}\stackrel{(b)}{=}\E_{|\tilde{h}_2|^2\!<\!{J^{(1)}_{\rm th}}\big(\boldsymbol{\mathcal{I}}_1\!+\!C_{\mathcal{D}}/(a_1L_1P\tilde{g}^2)\big)}\!\!\left[\exp\left(-|\tilde{h}_2|^2\!\times\!10^{-s/10}\right)-\exp\left(-\gamma^{(1)}_{\rm th}\!\Big[|\tilde{h}_2|^2\!\times\!10^{-s/10}+{\boldsymbol{\mathcal{I}}_1}+C_{\mathcal{D}}/(a_1L_1P\tilde{g}^2)\Big]\right)\right]\nonumber\\
	&\hspace{-3.8cm}=\E_{\boldsymbol{\mathcal{I}}_1,\tilde{g}}\left[({1+10^{-s/10}})^{-1}\times\left[1-\exp\left(-\left[{1+10^{-s/10}}\right]\!{J^{(1)}_{\rm th}}\big(\boldsymbol{\mathcal{I}}_1\!+\!C_{\mathcal{D}}/(a_1L_1P\tilde{g}^2)\big)\right)\right]\right]-\!\E_{\boldsymbol{\mathcal{I}}_1,\tilde{g}}\!\bigg[({1\!+\!\gamma^{(1)}_{\rm th}\!\!\times\!10^{-s/10}})^{-1}\nonumber\\
	&\hspace{-3.5cm}\times\!\left[1\!-\!\exp\!\left(\!-\!\left[{1\!+\!\gamma^{(1)}_{\rm th}\!\!\times\!\!10^{-s/10}}\right]\!{J^{(1)}_{\rm th}}\big(\boldsymbol{\mathcal{I}}_1\!+\!C_{\mathcal{D}}/(a_1\!L_1\!P\tilde{g}^2)\big)\right)\right]\!\exp\!\left(\!-\gamma^{(1)}_{\rm th}\!\Big[{\boldsymbol{\mathcal{I}}_1}\!+\!C_{\mathcal{D}}/(a_1\!L_1\!P\tilde{g}^2)\Big]\!\right)\!\!\bigg]\nonumber\\
	&\hspace{-4.1cm}\stackrel{(c)}{=}\frac{10^{s/10}}{1+10^{s/10}}\left(1\!-\!\exp\!\bigg(\frac{-J^{(1)}_{{\rm th},1}\sigma^2_{\mathcal{R}}}{a_1L_1P}\bigg)\E_{\tilde{g}}\!\!\left[\exp\!\bigg(\frac{-J^{(1)}_{{\rm th},1}C_{\mathcal{D}}}{a_1L_1P\tilde{g}^2}\bigg)\right]\prod_{k=1}^{K}\frac{a_1L_1P}{a_1L_1P+{J^{(1)}_{{\rm th},1}}L'_kp'_k}\right)\!-\!\frac{10^{s/10}}{\gamma^{(1)}_{\rm th}+10^{s/10}}\Bigg(\!\!\exp\!\bigg(\frac{-\gamma^{(1)}_{\rm th}\sigma^2_{\mathcal{R}}}{a_1L_1P}\bigg)\nonumber\\
	&\hspace{-4.2cm}\times\!\E_{\tilde{g}}\!\!\left[\exp\!\bigg(\frac{-\gamma^{(1)}_{\rm th}C_{\mathcal{D}}}{a_1L_1P\tilde{g}^2}\bigg)\right]\prod_{k=1}^{K}\frac{a_1L_1P}{a_1L_1P+{\gamma^{(1)}_{\rm th}}L'_kp'_k}-\exp\bigg(\frac{-\sigma^2_{\mathcal{R}}J^{(1)}_{{\rm th},2}}{a_1L_1P}\bigg)\E_{\tilde{g}}\!\!\left[\exp\!\bigg(\frac{-C_{\mathcal{D}}J^{(1)}_{{\rm th},2}}{a_1L_1P\tilde{g}^2}\bigg)\right]\prod_{k=1}^{K}\frac{a_1L_1P}{a_1L_1P+L'_kp'_kJ^{(1)}_{{\rm th},2}}\Bigg).
	\end{align}
	\hrulefill
	\vspace{-0.25cm}
\end{figure*}

{Let $\gamma^{(i)}_{\rm th}=2^{R^{(i)}_{\rm th}}-1$ denote the threshold SINR for an IM/DD FSO link to achieve the desired data rate $R^{(i)}_{\rm th}$, $i=1,2$. Then the} outage probability of the first user $\mathcal{U}_1$ in achieving an individual rate of $R^{(1)}_{\rm th}$ can be characterized as
	\setcounter{equation}{5}\begin{align}\label{p_out_1}
\!\!P_{\rm out}^{(1)}&\stackrel{(a)}{=}P(\pi_1)P_{{\rm out}|\pi_1}^{(1)}+P(\pi_2)P_{{\rm out}|\pi_2}^{(1)}\nonumber\\
&\stackrel{(b)}{=} 1\!-\!\Big[\Pr(\gamma^{(1)}_{\pi_1}>\gamma^{(1)}_{\rm th},\pi_1)+\nonumber\\
&\hspace{-0.3cm}\Pr(\gamma^{(2)}_{\pi_2}>\gamma^{(2)}_{\rm th},\pi_2)\times\Pr(\gamma^{(1)}_{\pi_2}>\gamma^{(1)}_{\rm th},\pi_2)/P(\pi_2)\Big],\!
\end{align}
where step $(a)$ follows from the law of total probability by defining  $P_{{\rm out}|\pi_i}^{(1)}$, $i=1,2$, as the conditional outage probability of the first NOMA user given the decoding order $\pi_i$. Moreover,
\begin{align}\label{[ppi1]}
P(\pi_1)&=\Pr(|\tilde{h}_1|^2\geq |\tilde{h}_2|^2\times10^{-s/10})\nonumber\\
&=\E_{|\tilde{h}_2|^2}[\exp(-|\tilde{h}_2|^2\times10^{-s/10})]\nonumber\\
&=(1+10^{-s/10})^{-1},
\end{align}
 and $P(\pi_2)=1-P(\pi_1)=(1+10^{s/10})^{-1}$ are the probabilities of having decoding orders $\pi_1$ and $\pi_2$, respectively. Furthermore, step $(b)$ follows, first, by defining $P_{{\rm cov}|\pi_i}^{(1)}\triangleq1-P_{{\rm out}|\pi_i}^{(1)}$, $i=1,2$, as the probability of successfully achieving $R^{(1)}_{\rm th}$ for $\mathcal{U}_1$  conditioned on the decoding order $\pi_i$, and then noting that the correct detection of $x_1$ for the decoding order
 $\pi_2$ 
  also requires successful decoding of the preceding symbol $x_2$, i.e.,\footnote{More precisely, $P_{{\rm cov}|\pi_2}^{(1)}$ in \eqref{eq9} should be written in the form of the joint probability $P_{{\rm cov}|\pi_2}^{(1)}=\Pr(\gamma^{(2)}_{\pi_2}>\gamma^{(2)}_{\rm th},\gamma^{(1)}_{\pi_2}>\gamma^{(1)}_{\rm th}|\pi_2)=\Pr(\gamma^{(2)}_{\pi_2}>\gamma^{(2)}_{\rm th}|\pi_2)\times\Pr(\gamma^{(1)}_{\pi_2}>\gamma^{(1)}_{\rm th}|\gamma^{(2)}_{\pi_2}>\gamma^{(2)}_{\rm th},\pi_2)$ since the events $\{\gamma^{(1)}_{\pi_2}>\gamma^{(1)}_{\rm th}\}$ and $\{\gamma^{(2)}_{\pi_2}>\gamma^{(2)}_{\rm th}\}$ are not independent due to the presence of multiuser interference and the backhaul link imposing common random variables $\boldsymbol{\mathcal{I}}_1$ (defined in \eqref{Joint1,1}) and $\tilde{g}$ on both $\gamma^{(1)}_{\pi_2}$ and $\gamma^{(2)}_{\pi_2}$. However, these two events are independent conditioned on $\boldsymbol{\mathcal{I}}_1$ and $\tilde{g}$. Therefore, we can calculate $\Pr(\gamma^{(1)}_{\pi_2}>\gamma^{(1)}_{\rm th}|\gamma^{(2)}_{\pi_2}>\gamma^{(2)}_{\rm th},\pi_2)=\E_{\boldsymbol{\mathcal{I}}_1,\tilde{g}}\left[\Pr(\gamma^{(1)}_{\pi_2}>\gamma^{(1)}_{\rm th}|\gamma^{(2)}_{\pi_2}>\gamma^{(2)}_{\rm th},\pi_2,\boldsymbol{\mathcal{I}}_1,\tilde{g})\right]=\E_{\boldsymbol{\mathcal{I}}_1,\tilde{g}}\left[\Pr(\gamma^{(1)}_{\pi_2}>\gamma^{(1)}_{\rm th}|\pi_2,\boldsymbol{\mathcal{I}}_1,\tilde{g})\right]=\Pr(\gamma^{(1)}_{\pi_2}>\gamma^{(1)}_{\rm th}|\pi_2)$.}
 \begin{align}\label{eq9}
 P_{{\rm cov}|\pi_1}^{(1)}&=\Pr(\gamma^{(1)}_{\pi_1}>\gamma^{(1)}_{\rm th}|\pi_1),\nonumber\\
 P_{{\rm cov}|\pi_2}^{(1)}&=\Pr(\gamma^{(2)}_{\pi_2}>\gamma^{(2)}_{\rm th}|\pi_2)\times\Pr(\gamma^{(1)}_{\pi_2}>\gamma^{(1)}_{\rm th}|\pi_2).
 \end{align}
 In the following, we calculate the three joint probabilities in \eqref{p_out_1} to ascertain the outage probability of the first user $\mathcal{U}_1$. 

In order to calculate $\Pr(\gamma^{(1)}_{\pi_1}>\gamma^{(1)}_{\rm th},\pi_1)$ we first note that $\Pr(\gamma^{(1)}_{\pi_1}>\gamma^{(1)}_{\rm th},\pi_1)=P(\pi_1)\Pr(\gamma^{(1)}_{\pi_1}>\gamma^{(1)}_{\rm th}|\pi_1)=P(\pi_1)[1-\Pr(\gamma^{(1)}_{\pi_1}<\gamma^{(1)}_{\rm th}|\pi_1)]=P(\pi_1)-\Pr(\gamma^{(1)}_{\pi_1}<\gamma^{(1)}_{\rm th},\pi_1)$.
Then using \eqref{gamma_f}, $\Pr(\gamma^{(1)}_{\pi_1}<\gamma^{(1)}_{\rm th},\pi_1)$ {for $\gamma^{(1)}_{\rm th}<1$} can be calculated as \eqref{Joint1,1} shown at the top of this page where, in step $(a)$, ${\boldsymbol{\mathcal{I}}_1}\triangleq(\sum_{k=1}^{K}{L'_kp'_k}|\tilde{h}'_k|^2+\sigma^2_{\mathcal{R}})/(a_1L_1P)$
 is the sum of the power of multiuser interference and noise, at the relay, normalized to the average power of the first NOMA user.
  Moreover, step $(b)$ follows, first, by defining the constant $J^{(1)}_{\rm th}\triangleq10^{s/10}\times\gamma^{(1)}_{\rm th}/(1-\gamma^{(1)}_{\rm th})>0$ for $\gamma^{(1)}_{\rm th}<1$, and then noting that $\Pr(X<Y,X\geq Z)$ for three random variables (RVs) $X$, $Y$, and $Z$ can be calculated using the law of total probability as $\Pr(X<Y,X\geq Z)=\Pr(Z\leq X<Y, Z<Y)$ since $\Pr(Z\leq X<Y, Z\geq Y)=0$. Finally, step $(c)$ of \eqref{Joint1,1} follows by noting that for any constant $C$
\begin{align}\label{E_1,1}
\setcounter{equation}{9}
&\hspace{-0.1cm}\E_{\boldsymbol{\mathcal{I}}_1,\tilde{g}}\left[\exp\left(-C\big[\boldsymbol{\mathcal{I}}_1+C_{\mathcal{D}}/(a_1L_1P\tilde{g}^2)\big]\right)\right]=\nonumber\\
&\hspace{-0.2cm}\!\exp\!\bigg(\!\frac{-C\sigma^2_{\mathcal{R}}}{a_1L_1P}\!\bigg)\E_{\tilde{g}}\!\!\left[\exp\!\bigg(\!\frac{-CC_{\mathcal{D}}}{a_1\!L_1\!P\tilde{g}^2}\!\!\bigg)\!\right]\prod_{k=1}^{K}\!\frac{a_1L_1P}{a_1\!L_1\!P\!+\!C\!L'_kp'_k},\!
\end{align}
due to the independence of ${\boldsymbol{\mathcal{I}}}_1$ and $\tilde{g}$, and then applying the independency among $|\tilde{h}'_k|^2$'s to get $\E_{\boldsymbol{\mathcal{I}}_1}\!\big[\exp(-C\boldsymbol{\mathcal{I}}_1)\big]=\exp\big(\frac{-C\sigma^2_{\mathcal{R}}}{a_1L_1P}\big)\prod_{k=1}^{K}\E_{|{\tilde{h}'}_{k}|^2}\left[\exp\left(-CL'_kp'_k|\tilde{h}'_k|^2/(a_1L_1P)\right)\right]$.
Furthermore, in step $(c)$ of \eqref{Joint1,1}, $J^{(1)}_{{\rm th},1}\triangleq J^{(1)}_{\rm th}({1+10^{-s/10}})$ and $J^{(1)}_{{\rm th},2}\triangleq \gamma^{(1)}_{\rm th}+J^{(1)}_{\rm th}({1+\gamma^{(1)}_{\rm th}\times10^{-s/10}})$.

We should further emphasize that \eqref{Joint1,1} is obtained for $\gamma^{(1)}_{\rm th}<1$. If $\gamma^{(1)}_{\rm th}\geq1$, the upper limit of $|\tilde{h}_1|^2$ in the equality $(a)$ of \eqref{Joint1,1} is always greater than its lower limit meaning that the condition $(1-\gamma^{(1)}_{\rm th})|\tilde{h}_2|^2\times10^{-s/10}<\gamma^{(1)}_{\rm th}\left[{\boldsymbol{\mathcal{I}}_1}+C_{\mathcal{D}}/(a_1L_1P\tilde{g}^2)\right]$ holds for all values of $|\tilde{h}_2|^2$ and there is no need to impose such an extra condition on the derivation of the corresponding probability.
{Consequently, by averaging over $|\tilde{h}_2|^2$, $\boldsymbol{\mathcal{I}}_1$, and $\tilde{g}$, $\Pr(\gamma^{(1)}_{\pi_1}<\gamma^{(1)}_{\rm th},\pi_1)$ for $\gamma^{(1)}_{\rm th}\geq1$ can be derived as \eqref{E1,3} shown at the top of the next page.}

\begin{figure*}[!t]
	\normalsize
	\vspace{-0.25cm}
	\begin{align}\label{E1,3}
	\Pr(\gamma^{(1)}_{\pi_1}<\gamma^{(1)}_{\rm th},\pi_1){\Big |}_{\gamma^{(1)}_{\rm th}\geq1}\!\!{=}\frac{10^{s/10}}{1\!+\!10^{s/10}}-\frac{10^{s/10}}{\gamma^{(1)}_{\rm th}\!+\!10^{s/10}}\exp\!\bigg(\!\frac{-\gamma^{(1)}_{\rm th}\sigma^2_{\mathcal{R}}}{a_1L_1P}\!\bigg)\!\E_{\tilde{g}}\!\!\left[\exp\!\bigg(\frac{-\gamma^{(1)}_{\rm th}C_{\mathcal{D}}}{a_1L_1P\tilde{g}^2}\bigg)\right]\prod_{k=1}^{K}\frac{a_1L_1P}{a_1L_1P\!+\!{\gamma^{(1)}_{\rm th}}L'_kp'_k}.
	\end{align}
	\hrulefill
	\vspace{-0.25cm}
\end{figure*}
\begin{figure*}[!t]
	\normalsize
	\vspace{-0.25cm}
	\begin{align}\label{E_g}
	\mathcal{G}(A)\!\triangleq\!\E_{\tilde{g}}\!\left[\exp\!\left(\!-\frac{A}{\tilde{g}^2}\right)\!\right]\!\!=\!\frac{\xi^2\!\times\!2^{\alpha+\beta-2}}{2\pi \Gamma(\alpha)\Gamma(\beta)}{\rm G}_{6,1}^{0,6}\left[\frac{16A_0^2}{A(\alpha\beta)^2}{\bigg |\begin{matrix}
		1,(2-\xi^2)/2,(1-\alpha)/2,(2-\alpha)/2,(1-\beta)/2,(2-\beta)/2\\ -\xi^2/2
		\end{matrix}}\right].
	\end{align}
	\vspace{-0.25cm}
	\hrulefill
\end{figure*}
\begin{figure*}[!t]
	\normalsize
	\vspace{-0.25cm}
	\begin{align}\label{E2,1}
	\Pr(\gamma^{(2)}_{\pi_2}<\gamma^{(2)}_{\rm th},\pi_2){\Big|}_{\gamma^{(2)}_{\rm th}<1}\!&=\frac{1}{1\!+\!10^{s/10}}\Bigg(\!1\!-\!\exp\!\bigg(\frac{-J^{(2)}_{{\rm th},1}\sigma^2_{\mathcal{R}}}{a_2L_2P}\bigg)\mathcal{G}\!\bigg(\frac{J^{(2)}_{{\rm th},1}C_{\mathcal{D}}}{a_2L_2P}\bigg)\!\prod_{k=1}^{K}\frac{a_2L_2P}{a_2L_2P\!+\!{J^{(2)}_{{\rm th},1}}L'_kp'_k}\!\Bigg)-\frac{10^{-s/10}}{\gamma^{(2)}_{\rm th}+10^{-s/10}}\nonumber\\
	&\hspace{-4cm}\times\Bigg[\exp\!\bigg(\frac{-\gamma^{(2)}_{\rm th}\sigma^2_{\mathcal{R}}}{a_2L_2P}\bigg)\mathcal{G}\!\bigg(\frac{\gamma^{(2)}_{\rm th}C_{\mathcal{D}}}{a_2L_2P}\bigg)\prod_{k=1}^{K}\frac{a_2L_2P}{a_2L_2P+{\gamma^{(2)}_{\rm th}}L'_kp'_k}-\exp\!\bigg(\frac{-\sigma^2_{\mathcal{R}}J^{(2)}_{{\rm th},2}}{a_2L_2P}\bigg)\mathcal{G}\!\bigg(\frac{C_{\mathcal{D}}J^{(2)}_{{\rm th},2}}{a_2L_2P}\bigg)\prod_{k=1}^{K}\frac{a_2L_2P}{a_2L_2P+L'_kp'_kJ^{(2)}_{{\rm th},2}}\Bigg].
	\end{align}
	\hrulefill
	\vspace{-0.25cm}
\end{figure*}
\begin{figure*}[!t]
	\normalsize
	\vspace{-0.25cm}
	\begin{align}\label{E2,2}
	\Pr(\gamma^{(2)}_{\pi_2}<\gamma^{(2)}_{\rm th},\pi_2){\Big |}_{\gamma^{(2)}_{\rm th}\geq1}\!\!{=}\frac{1}{1\!+\!10^{s/10}}-\frac{10^{-s/10}}{\gamma^{(2)}_{\rm th}\!+\!10^{-s/10}}\exp\!\bigg(\!\frac{-\gamma^{(2)}_{\rm th}\sigma^2_{\mathcal{R}}}{a_2L_2P}\!\bigg)\mathcal{G}\!\bigg(\frac{\gamma^{(2)}_{\rm th}C_{\mathcal{D}}}{a_2L_2P}\bigg)\prod_{k=1}^{K}\frac{a_2L_2P}{a_2L_2P\!+\!{\gamma^{(2)}_{\rm th}}L'_kp'_k}.
	\end{align}
	\hrulefill
	\vspace{-0.25cm}
\end{figure*}

Finally, using \eqref{Joint1,1} for $\gamma^{(1)}_{\rm th}<1$ or \eqref{E1,3} for $\gamma^{(1)}_{\rm th}\geq1$, one can obtain the coverage probability of the first NOMA user for the decoding order $\pi_1$ as $P_{\rm cov}^{(1)}(\pi_1)=\Pr(\gamma^{(1)}_{\pi_1}>\gamma^{(1)}_{\rm th},\pi_1)=(1+10^{-s/10})^{-1}-\Pr(\gamma^{(1)}_{\pi_1}<\gamma^{(1)}_{\rm th},\pi_1)$. However, the closed-form characterization of $P_{\rm cov}^{(1)}(\pi_1)$ still requires the calculation of expressions of the form $\E_{\tilde{g}}\!\left[\exp\left(-A/\tilde{g}^2\right)\right]$, where $A$ is a constant and $\tilde{g}$ is distributed according to \eqref{f_gtild}. To do so, we first apply \cite[Eq. (11)]{adamchik1990algorithm} and \cite[Eq. (9.31.2)]{gradshteyn2014table} to write $\exp\left(-A/\tilde{g}^2\right)$ in the form of a Meijer's G-function as $\exp\left(-A/\tilde{g}^2\right)={\rm G}_{1,0}^{0,1}\left[\tilde{g}^2/A\big |^1_{-}\right]$. Then we can apply \cite[Eq. (21)]{adamchik1990algorithm} to calculate the infinite integral of product of Meijer's G-functions involved in $\E_{\tilde{g}}\!\!\left[\exp\left(-A/\tilde{g}^2\right)\right]=\int_{0}^{\infty}\exp\left(-A/\tilde{g}^2\right)f_{\tilde{g}}(\tilde{g})d\tilde{g}$ as \eqref{E_g} shown at the top of the next page. { Note that the order of G-function in \eqref{E_g} is reduced using \cite[Eq. (9.31.1)]{gradshteyn2014table}.} For the ease of notation, hereafter, we denote $\E_{\tilde{g}}\!\!\left[\exp\left(-A/\tilde{g}^2\right)\right]$ by $\mathcal{G}(A)$ for any constant $A$.

Similarly, the second term in \eqref{p_out_1} can be obtained, first, by writing $\Pr(\gamma^{(2)}_{\pi_2}>\gamma^{(2)}_{\rm th},\pi_2)=P(\pi_2)-\Pr(\gamma^{(2)}_{\pi_2}<\gamma^{(2)}_{\rm th},\pi_2)$. Then using the symmetry of the problem, it can be shown that $\Pr(\gamma^{(2)}_{\pi_2}<\gamma^{(2)}_{\rm th},\pi_2)$ for $\gamma^{(2)}_{\rm th}<1$ and $\gamma^{(2)}_{\rm th}\geq 1$ can be obtained as \eqref{E2,1} and \eqref{E2,2}, respectively, shown at the top of the next page, where $\mathcal{G}(\cdot)$ is given in \eqref{E_g}, and $J^{(2)}_{\rm th}\triangleq 10^{-s/10}\times\gamma^{(2)}_{\rm th}/(1-\gamma^{(2)}_{\rm th})>0$ is defined for $\gamma^{(2)}_{\rm th}<1$. Also, $J^{(2)}_{{\rm th},1}\triangleq J^{(2)}_{\rm th}({1+10^{s/10}})$ and $J^{(2)}_{{\rm th},2}\triangleq \gamma^{(2)}_{\rm th}+J^{(2)}_{\rm th}({1+\gamma^{(2)}_{\rm th}\times10^{s/10}})$.
\begin{figure*}[!t]
	\normalsize
	\vspace{-0.25cm}
	\begin{align}\label{P_1_3}
	\!\!\!\!\!\Pr(\gamma^{(1)}_{\pi_2}\!>\!\gamma^{(1)}_{\rm th}\!,\pi_2)\!=\!\frac{1}{1\!+\!10^{s/10}}\!\times\!\exp\!\bigg(\!\!\frac{-\gamma^{(1)}_{\rm th}\sigma^2_{\mathcal{R}}(1\!+\!10^{s/10})}{a_1L_1P}\bigg)\mathcal{G}\!\bigg(\frac{\gamma^{(1)}_{\rm th}C_{\mathcal{D}}(1\!+\!10^{s/10})}{a_1L_1P}\bigg)\!\!\prod_{k=1}^{K}\!\frac{a_1L_1P}{a_1L_1P+{\gamma^{(1)}_{\rm th}}L'_kp'_k(1\!+\!10^{s/10})}.
	\end{align}
	\vspace{-0.25cm}
	\hrulefill
\end{figure*}

Moreover, the last probability term in \eqref{p_out_1} can be calculated by first writing $\Pr(\gamma^{(1)}_{\pi_2}>\gamma^{(1)}_{\rm th},\pi_2)=\Pr\left(\gamma^{(1)}_{\rm th}\left[{\boldsymbol{\mathcal{I}}_1}+C_{\mathcal{D}}/(a_1L_1P\tilde{g}^2)\right]<|\tilde{h}_1|^2\!\!<|\tilde{h}_2|^2\times10^{-s/10}\right)$, { where $\gamma^{(1)}_{\pi_2}$ can be expressed similar to \eqref{gamma_s}}. Then using a similar approach to \eqref{Joint1,1}, the closed-form expression for all values of $\gamma^{(1)}_{\rm th}$ can be expressed as \eqref{P_1_3} shown at the top of this page. This completes the closed-form characterization of the outage probability of the first NOMA user $\mathcal{U}_1$.

Finally, the outage probability of the second NOMA user $\mathcal{U}_2$ can be characterized as
\begin{align}\label{p_out_2}
\!\!P_{\rm out}^{(2)}&= 1-\Big[\Pr(\gamma^{(2)}_{\pi_2}>\gamma^{(2)}_{\rm th},\pi_2)+\nonumber\\
&\hspace{-0.5cm}\Pr(\gamma^{(1)}_{\pi_1}>\gamma^{(1)}_{\rm th},\pi_1)\times\Pr(\gamma^{(2)}_{\pi_1}>\gamma^{(2)}_{\rm th},\pi_1)/P(\pi_1)\Big],
\end{align}
where $\Pr(\gamma^{(2)}_{\pi_2}>\gamma^{(2)}_{\rm th},\pi_2)$ and $\Pr(\gamma^{(1)}_{\pi_1}>\gamma^{(1)}_{\rm th},\pi_1)$ have already been calculated, and, { using the symmetry of the problem}, $\Pr(\gamma^{(2)}_{\pi_1}>\gamma^{(2)}_{\rm th},\pi_1)$ can be obtained as
\begin{align}\label{P_out_2,2}
&\Pr(\gamma^{(2)}_{\pi_1}>\gamma^{(2)}_{\rm th},\pi_1)=(1+10^{-s/10})^{-1}\times\nonumber\\
&\exp\!\bigg(\frac{-\gamma^{(2)}_{\rm th}\sigma^2_{\mathcal{R}}(1\!+\!10^{-s/10})}{a_2L_2P}\bigg)\mathcal{G}\!\bigg(\frac{\gamma^{(2)}_{\rm th}C_{\mathcal{D}}(1\!+\!10^{-s/10})}{a_2L_2P}\bigg)\nonumber\\
&\hspace{1.4cm}\times\prod_{k=1}^{K}\!\frac{a_2L_2P}{a_2L_2P+{\gamma^{(2)}_{\rm th}}L'_kp'_k(1+10^{-s/10})}.
\end{align}

{\textbf{Remark 1:} The latter analysis suggests that the outage probability of the second NOMA user can be characterized using the preceding analysis by substituting $-s$ for $s$ and appropriate change of indexing $1 \leftrightarrow 2$. This is because the only difference between $\mathcal{U}_1$ and $\mathcal{U}_2$ is that the user with a lower average gain is labeled as the second user, i.e., $a_2L_2=a_1L_1\times10^{-s/10}$.}
\subsection{Sum-Rate Outage Analysis}
 For the uplink NOMA transmission with logarithmic functions for the rates as $R^{(i)}_{\pi_j}=\log_2(1+\gamma^{(i)}_{\pi_j})$, $i,j\in\{1,2\}$, it can be verified that the sum  of the NOMA users, regardless of their decoding order, can be expressed as
 {
\begin{align}\label{R_sum}
R_{\Sigma}\!=&\log_2([1+\gamma^{(1)}_{\pi_j}][1+\gamma^{(2)}_{\pi_j}])\nonumber\\
=&\log_2\!\left(\!1\!+\!\frac{a_1L_1P\tilde{g}^2|\tilde{h}_1|^2+a_2L_2P\tilde{g}^2|\tilde{h}_2|^2}{\sum_{k=1}^{K}\!{L'_kp'_k}\tilde{g}^2|\tilde{h}'_k|^2+\tilde{g}^2\sigma^2_{\mathcal{R}}\!+\!C_{\mathcal{D}}}\!\right).
\end{align}
}
Denoting the fractional term of the logarithm argument in \eqref{R_sum} by $\gamma_{\Sigma}$, the sum-rate outage probability defined as $P_{\rm out}^{\Sigma}=\Pr(\gamma_{\Sigma}<\gamma^{\Sigma}_{\rm th})$, where $\gamma^{\Sigma}_{\rm th}=2^{R^{\Sigma}_{\rm th}}-1$ is the threshold equivalent SINR to achieve the desired sum rate of $R^{\Sigma}_{\rm th}$, can be expressed as
\begin{align}\label{P_out_sum}
\!\!\!\!P_{\rm out}^{\Sigma}\!=\!\Pr\!\left(\!|\tilde{h}_1|^2\!<\!\gamma^{\Sigma}_{\rm th}\!\!\left[\boldsymbol{\mathcal{I}}_1\!\!+\!\frac{C_{\mathcal{D}}}{a_1\!L_1\!P\tilde{g}^2}\right]\!\!-\!|\tilde{h}_2|^2\!\times\!10^{-s/10}\!\right)\!.\!
\end{align}

Let $\mathcal{B}$ represent the event $\{|\tilde{h}_2|^2<\gamma^{\Sigma}_{\rm th}\!\times\!10^{s/10}[\boldsymbol{\mathcal{I}}_1\!+\!C_{\mathcal{D}}\!/\!(a_1\!L_1\!P\tilde{g}^2)]\}$, and $\mathcal{O_S}\triangleq\{|\tilde{h}_1|^2\!<\!\gamma^{\Sigma}_{\rm th}\!\left[\boldsymbol{\mathcal{I}}_1\!\!+\!C_{\mathcal{D}}\!/\!(a_1\!L_1\!P\tilde{g}^2)\right]\!\!-\!|\tilde{h}_2|^2\!\times\!10^{-s/10}\}$ denote the sum-rate outage event defined in \eqref{P_out_sum}. Clearly, $\Pr(\mathcal{O_S},\mathcal{B}^c)=0$ where $\mathcal{B}^c$ is the complementary event of $\mathcal{B}$. Therefore, using the law of total probability, $P_{\rm out}^{\Sigma}$ can be expressed as $P_{\rm out}^{\Sigma}=\Pr(\mathcal{O_S},\mathcal{B})$, which is calculated in a closed form as \eqref{Pout_sum} at the top of the next page. 
\begin{figure*}[!t]
	\normalsize
	\vspace{-0.25cm}
	\begin{align}\label{Pout_sum}
	P_{\rm out}^{\Sigma}=&\E_{|\tilde{h}_2|^2<\gamma^{\Sigma}_{\rm th}\!\times\!{10^{s/10}}\big[\boldsymbol{\mathcal{I}}_1+C_{\mathcal{D}}/(a_1L_1P\tilde{g}^2)\big]}\left[1-\exp\left(-\gamma^{\Sigma}_{\rm th}\!\left[\boldsymbol{\mathcal{I}}_1+C_{\mathcal{D}}/(a_1L_1P\tilde{g}^2)\right]+|\tilde{h}_2|^2\times10^{-s/10}\right)\right]\nonumber\\
	=&1+\frac{1}{10^{s/10}-1}\times\exp\!\bigg(\frac{-\sigma^2_{\mathcal{R}}\gamma^{\Sigma}_{\rm th}\times10^{s/10}}{a_1L_1P}\bigg){\mathcal{G}}\!\bigg(\frac{C_{\mathcal{D}}\gamma^{\Sigma}_{\rm th}\times10^{s/10}}{a_1L_1P}\bigg)\prod_{k=1}^{K}\frac{a_1L_1P}{a_1L_1P+L'_kp'_k\gamma^{\Sigma}_{\rm th}\times10^{s/10}}\nonumber\\
	&{\hspace{0.2cm}}-\frac{10^{s/10}}{10^{s/10}-1}\times\exp\!\bigg(\frac{-\gamma^{\Sigma}_{\rm th}\sigma^2_{\mathcal{R}}}{a_1L_1P}\bigg){\mathcal{G}}\!\bigg(\frac{\gamma^{\Sigma}_{\rm th}C_{\mathcal{D}}}{a_1L_1P}\bigg)\prod_{k=1}^{K}\frac{a_1L_1P}{a_1L_1P+\gamma^{\Sigma}_{\rm th}L'_kp'_k}.
	\end{align}
	\hrulefill
	\vspace{-0.25cm}
\end{figure*}

{\textbf{Remark 2:} In the special case of the absence of multiuser interference (except the NOMA users themselves), one can obtain the outage probability closed-form expressions by substituting $L'_kp'_k=0$, $\forall k=1,2,...,K$, which replaces all the product terms of the form $\prod_{k=1}^K[\cdot]$ by $1$ and summations of the form $\sum_{k=1}^K[\cdot]$ by $0$.}
\subsection{RF Backhaul Analysis}
{In this part, for the sake of completeness and comparison, we extend our preceding analysis to the case of conventional RF-backhauled systems. In particular,} we investigate the performance of dual-hop uplink NOMA where the $\mathcal{R-D}$ backhaul link forwards the amplified version of $y_{\mathcal{R}}$ in \eqref{y_R} by a gain $G_b$ through an RF link with the path-loss gain $L_b$ and fading coefficient $\tilde{h}_b$, i.e., with the composite channel gain $h_b=\sqrt{L_b}\tilde{h}_b$. For the backhaul link, a line-of-sight (LOS) path should be available from the relay to destination for the applicability of the directive FSO link.  
Therefore, for the $\mathcal{R-D}$ RF backhaul link we assume Rician fading, with the shape parameter $\Omega$ and scale parameter $\Psi$, which takes into account the effects of both LOS and scattered paths. The scale parameter $\Psi$ is the total average power of fading and hence $\Psi=1$, and $\Omega$ represents the ratio of the power contributions by the LOS path to the remaining scattered paths. In this case, $\kappa_b\triangleq|\tilde{h}_b|^2$ can be characterized according to a non-central chi-square distribution given by
\begin{align}\label{pdf_rice}
f_{\kappa_b}(\kappa_b)\!=\!\frac{1\!+\!\Omega}{{\rm e}^{\Omega}}\exp\!\left(\!-{(1\!+\!\Omega)\kappa_b}\right)I_0\!\left(2\sqrt{{\Omega(1\!+\!\Omega)\kappa_b}}\right)\!,\!
\end{align}
where $I_0(\cdot)$ is the zeroth-order modified Bessel function of the first kind \cite{simon2005digital}.

{Given that the second hop is employing the RF band, it is practically possible for the destination receiver to be exposed to the \textit{presence} of some external interfering users.
In what follows we characterize the outage probability of the uplink NOMA transmission when the destination, in addition to the directive signal from the $\mathcal{R-D}$ Rician channel, receives the superposition of the signals of $K_{\rm d}$ interfering users $\mathcal{I}''_j$, $j=1,2,...,K_{\rm d}$, each with the transmit power $p''_j$, path-loss gain $L''_j$, and i.i.d. Rayleigh fading coefficient $\tilde{h}''_j$.
In this case, the received signal can be expressed as $y^{\rm RF}_\mathcal{D}=\sqrt{L_b}\tilde{h}_bG_by_\mathcal{R}+\sum_{j=1}^{K_{\rm d}}x''_j\tilde{h}''_j\sqrt{L''_jp''_j}+n^{\rm RF}_\mathcal{D}$, where $x''_j$ is the transmitted signal by the $j$-th interfering users and $n^{\rm RF}_\mathcal{D}$ is the RF noise at the destination with mean zero and variance $N_0$. 
Therefore, it is easy to verify that all the SINR equations (e.g., \eqref{gamma_f} and \eqref{gamma_s}) remain the same except that $\tilde{g}^2$ and $C_{\mathcal{D}}$ will be replaced by $\kappa_b\triangleq|\tilde{h}_b|^2$ and $(N_0+\boldsymbol{\mathcal{I}}'')/(L_bG_b^2)=C^{\rm RF}_{\mathcal{D}}+\boldsymbol{\mathcal{I}}''/(L_bG_b^2)$, respectively, where $\boldsymbol{\mathcal{I}}''\triangleq\sum_{j=1}^{K_{\rm d}}{L''_jp''_j}|\tilde{h}''_j|^2$ is the total external interference to the destination and $C^{\rm RF}_{\mathcal{D}}\triangleq N_0/(L_bG_b^2)$. {Therefore, in order to characterize the outage probability formulas, instead of calculating $\E_{\tilde{g}}[\exp(-BC_{\mathcal{D}}/\tilde{g}^2)]$ or equivalently ${\mathcal{G}}(BC_{\mathcal{D}})$,
	we need to evaluate $\E_{\kappa_b,\boldsymbol{\mathcal{I}}''}[\exp(-B(\boldsymbol{\mathcal{I}}''+N_0)/(L_bG_b^2\kappa_b))]$. Note that the constant $B$ represents the factor of $-C_{\mathcal{D}}/\tilde{g}^2$ or $C_{\mathcal{D}}$ in the arguments of $\E_{\tilde{g}}[\exp(\cdot)]$ or ${\mathcal{G}}(\cdot)$, respectively, in the previously-derived formulas. The required expectation can be characterized as}
\begin{align}\label{E_I''}
&\E_{\kappa_b,\boldsymbol{\mathcal{I}}''}[\exp(-B(\boldsymbol{\mathcal{I}}''+N_0)/(L_bG_b^2\kappa_b))]\nonumber\\
&=\E_{\kappa_b}\!\!\left[\exp\!\left(\!-\frac{BC_{\mathcal{D}}^{\rm RF}}{\kappa_b}\right)\!\times\!\E_{\boldsymbol{\mathcal{I}}''}\!\!\Bigg[\prod_{j=1}^{K_{\rm d}}\exp\!\left(\!-\frac{B{L''_jp''_j}|\tilde{h}''_j|^2}{L_bG_b^2\kappa_b}\right)\!\Bigg]\!\right]\nonumber\\
&=\E_{\kappa_b}\!\!\left[\exp\!\left(\!-\frac{BC_{\mathcal{D}}^{\rm RF}}{\kappa_b}\right)\!\times\!\prod_{j=1}^{K_{\rm d}}\frac{L_bG_b^2\kappa_b}{L_bG_b^2\kappa_b+B{L''_jp''_j}}\right]\nonumber\\
&\stackrel{(a)}{=}\!\frac{1\!+\!\Omega}{{\rm e}^{\Omega}}\times\left(\frac{L_bG_b^2}{B}\right)^{\!\!K_{\rm d}-1}\sum_{l=1}^{K_{\rm d}}\Bigg(\prod_{\substack{j=1 \\ j\neq l}}^{K_{\rm d}}\frac{1}{L''_jp''_j-L''_lp''_l}\Bigg)\times\!\nonumber\\
&~~\int_{0}^{\infty}\frac{x^{K_{\rm d}}\!\times\!{\rm e}^{-(1+\Omega)x-BC_{\mathcal{D}}^{\rm RF}\!/x}}{x+BL''_lp''_l/(L_bG_b^2)}I_0\!\left(2\sqrt{{\Omega(1\!+\!\Omega)x}}\right)\!dx\nonumber\\
&{\stackrel{(b)}{=}\frac{1\!+\!\Omega}{{\rm e}^{\Omega}}\times\left(\frac{L_bG_b^2}{B}\right)^{\!\!K_{\rm d}-1}\sum_{l=1}^{K_{\rm d}}\Bigg\{\Bigg(\prod_{\substack{j=1 \\ j\neq l}}^{K_{\rm d}}\frac{1}{L''_jp''_j-L''_lp''_l}\Bigg)}\nonumber\\
&~~{\times\sum_{k=0}^{\infty}\Bigg\{\!\frac{(\Omega(1\!+\!\Omega))^k}{(k!)^2}{\left(\!\frac{BL''_lp''_l}{L_bG_b^2}\right)^{\!\!K_{\rm d}+k}\!\!\!\exp\!\left(\!\frac{BL''_lp''_l(1\!+\!\Omega)}{L_bG_b^2}\right)}}\nonumber\\
&~~{\times\mathsf{G}_{1,2}^{2,1}\!\left[\!\frac{L_bG_b^2C_{\mathcal{D}}^{\rm RF}}{L''_lp''_l}{\bigg |\begin{matrix}
		(K_{\rm d}\!+\!k\!+\!1,{BL''_lp''_l(1\!+\!\Omega)}/{(L_bG_b^2)})\\(0,0),(K_{\rm d}+k+1,0)
		\end{matrix}}\right]\!\!\Bigg\}\!\!\Bigg\},}
\end{align}
where $(a)$ is obtained by the partial fraction decomposition of the involved rational function assuming $L''_jp''_j\neq L''_lp''_l\neq0$, $\forall j\neq l$, which is a reasonable assumption given that $L''_jp''_j$'s are continuous quantities{\footnote{{From the stochastic geometry perspective, the interfering users are located in stochastic locations leading to path-losses that are continuous RVs. Hence, the probability of having \textit{exactly} equal $L''_kp''_k$'s in a \textit{real-world} scenario is zero. Accordingly, our derived closed-form expressions hold with probability one in real-world networks. However, if, \textit{hypothetically}, some users have equal distances and hence equal $L''_kp''_k$'s, then some of our derived formulas cannot further be simplified, e.g.,
we cannot write the equality in step $(a)$ of \eqref{E_I''}.}}}. 
{Moreover, step $(b)$ is obtained according to Appendix A, where $\mathsf{G}_{.,.}^{.,.}$ denotes the upper incomplete Meijer's G-function which is a special case of the generalized upper incomplete Fox's H-function $\mathcal{H}_{.,.}^{.,.}$ introduced in \cite{yilmaz2009product}.\footnote{A simple implementation of $\mathcal{H}_{.,.}^{.,.}$ in Mathematica  is presented in \cite[Appendix B]{yilmaz2009product}.}}

{\textbf{Remark 3:} As a special case, in the \textit{absence} of external RF interference to the destination, the received signal can be expressed as $y^{\rm RF}_\mathcal{D}=\sqrt{L_b}\tilde{h}_bG_by_\mathcal{R}+n^{\rm RF}_\mathcal{D}$; hence, $\boldsymbol{\mathcal{I}}''=0$. Therefore, all the preceding results in Sections III-A and III-B are valid for the RF-backhauled NOMA relaying system when $C_{\mathcal{D}}^{\rm RF}$ and $\E_{\kappa_b}[\exp(-A/\kappa_b)]$ are substituted for $C_{\mathcal{D}}$ and $\E_{\tilde{g}}[\exp(-A/\tilde{g}^2)]$, respectively, where $A$ is a constant, i.e., the argument of $\mathcal{G}(\cdot)$ in the formulas derived earlier.
	A particularly interesting form for $\E_{\kappa_b}[\exp(-A/\kappa_b)]$ can be obtained by applying \cite[Eq. (8.447.1)]{gradshteyn2014table} to expand $I_0\!\left(2\sqrt{{\Omega(1\!+\!\Omega)\kappa_b}}\right)$ in the form of an infinite series and then applying \cite[Eq. (3.471.9)]{gradshteyn2014table} to express $\E_{\kappa_b}[\exp(-A/\kappa_b)]$ as
	\begin{align}\label{E_kappa}
	&\!\E_{\kappa_b}[\exp(-A/\kappa_b)]=\nonumber\\
	&\hspace{0.3cm}\sum_{n=0}^{\infty}2{\rm e}^{-\Omega}[A(1+\Omega)]^{\frac{n+1}{2}}\frac{\Omega^n}{(n!)^2}K_{n+1}(2\sqrt{A(1+\Omega)}),
	\end{align}
	where $K_n(\cdot)$ is the $n$-th order modified Bessel function of the second kind. 
	As verified in \cite{suraweera2009two}, the summation in \eqref{E_kappa} can  effectively be calculated using only few terms, resulting in closed-form expressions for the outage probabilities.}}

Finally, it is worth mentioning that given any threshold/target data rate, the individual data rates of $R^{(i)}_{\rm th}(1-P_{\rm out}^{(i)})$, $i=1,2$, and sum rate of $R^{\Sigma}_{\rm th}(1-P_{\rm out}^{\Sigma})$ are achievable where the outage probabilities are calculated with respect to the threshold rates $R^{(1)}_{\rm th}$, $R^{(2)}_{\rm th}$, and $R^{\Sigma}_{\rm th}$ according to the analysis in this section. In addition to this characterization of achievable rate regions, we are also interested to determine the average data rates of the users, known as their \textit{ergodic capacity}, which is obtained using the assumption that the users' threshold/target data rates are adjusted by their channel conditions, i.e., $\gamma^{(i)}_{\rm th}=\gamma^{(i)}$ and $\gamma^{\Sigma}_{\rm th}=\gamma_{\Sigma}$, meaning that they can always decode their signals without outage. 
Such analysis will be proceeded in the next section in a same sequential order as this section.

\section{Ergodic Capacity Analysis}
\subsection{Average Individual Rates}
Given that the instantaneous individual rate $R_i$ of the $i$-th NOMA user, $i=1,2$, is related to its instantaneous SINR $\gamma^{(i)}$ as $R_i=\log_2(1+\gamma^{(i)})$, and that $\gamma^{(i)}$ can take two different forms of $\gamma^{(i)}_{\pi_1}$ and $\gamma^{(i)}_{\pi_2}$ each with the probabilities $P(\pi_1)$ and $P(\pi_2)$, respectively, the average individual rate $\overline{R}_i$ of the $i$-th NOMA user can be characterized as
\begin{align}\label{R_ibar}
\overline{R}_i&=\E_{\boldsymbol{\mathcal{I}}_1,\tilde{g},|\tilde{h}_2|^2,|\tilde{h}_1|^2>|\tilde{h}_2|^2\times10^{-s/10}}\left[\log_2(1+\gamma^{(i)}_{\pi_1})\right]\nonumber\\
&+\E_{\boldsymbol{\mathcal{I}}_1,\tilde{g},|\tilde{h}_2|^2,|\tilde{h}_1|^2<|\tilde{h}_2|^2\times10^{-s/10}}\left[\log_2(1+\gamma^{(i)}_{\pi_2})\right].
\end{align}
In this subsection, we first characterize the first user's average rate $\overline{R}_1$ (equivalently, calculate the two expectation terms in \eqref{R_ibar} for $i=1$), and then relate $\overline{R}_2$ to $\overline{R}_1$.
\begin{figure*}[!t]
	\normalsize
	\vspace{-0.25cm}
	\begin{align}\label{E_f1}
	&\E_{\boldsymbol{\mathcal{I}}_1,\tilde{g},|\tilde{h}_2|^2,|\tilde{h}_1|^2>|\tilde{h}_2|^2\times10^{-s/10}}\!\!\left[\log_2(1\!+\!\gamma^{(1)}_{\pi_1})\!\right]\!\!
	\stackrel{(a)}{=}\!\E_{\boldsymbol{\mathcal{I}}_1,\tilde{g},|\tilde{h}_2|^2}\!\!\left[\!\int_{\!|\tilde{h}_2|^2\!\times\!10^{-s\!/\!10}}^{\infty}\!\!\!\!\log_2\!\!\left(\!1\!+\!\frac{x}{|\tilde{h}_2|^2\!\!\times\!\!10^{-s\!/\!10}\!+\!{\boldsymbol{\mathcal{I}}_1}\!\!+\!C_{\!\mathcal{D}}\!/\!(a_1\!L_1\!P\tilde{g}^2)}\!\right)\!\!f_{|\tilde{h}_1|^2}(x)dx\!\right]\nonumber\\
	&\hspace{1cm}\stackrel{(b)}{=}\frac{1}{\ln 2}\underbrace{\E_{\boldsymbol{\mathcal{I}}_1,\tilde{g},|\tilde{h}_2|^2}\!\left[\exp\!\left(\!-|\tilde{h}_2|^2\!\times\!10^{-s/10}\right)\times\ln\!\left(\!1+\frac{|\tilde{h}_2|^2\!\times\!10^{-s/10}}{|\tilde{h}_2|^2\!\times\!10^{-s/10}+{\boldsymbol{\mathcal{I}}_1}+C_{\mathcal{D}}/(a_1L_1P\tilde{g}^2)}\right)\right]}_{V_1}\nonumber\\
	&\hspace{4.5cm}+\frac{1}{\ln 2}\underbrace{\E_{\boldsymbol{\mathcal{I}}_1,\tilde{g},|\tilde{h}_2|^2}\!\left[\int_{|\tilde{h}_2|^2\times10^{-s/10}}^{\infty}\frac{\exp(-x)}{x+|\tilde{h}_2|^2\!\times\!10^{-s/10}+{\boldsymbol{\mathcal{I}}_1}+C_{\mathcal{D}}/(a_1L_1P\tilde{g}^2)}dx\right]}_{V_2}.
	\end{align}
	\hrulefill
	\vspace{-0.25cm}
\end{figure*}
\begin{figure*}[!t]
		\setcounter{equation}{27}
	\normalsize
	\vspace{-0.25cm}
	\begin{align}\label{V1_avgI}
	&\E_{\boldsymbol{\mathcal{I}}_1,\tilde{g}}\!\Bigg[\exp\!\left(\!v_{1,i_1}\bigg[{\boldsymbol{\mathcal{I}}_1}+\frac{C_{\mathcal{D}}}{a_1L_1P\tilde{g}^2}\bigg]\right){\rm Ei}\!\left(-v_{1,i_1}\!\bigg[{\boldsymbol{\mathcal{I}}_1}+\frac{C_{\mathcal{D}}}{a_1L_1P\tilde{g}^2}\bigg]\right)\!\Bigg]=\nonumber\\
	&\!\beta^{(K)}_{v_{i_1}}\E_{\tilde{g}}\!\left[\exp\!\left(\!v_{1,i_1}\frac{\sigma^2_{\mathcal{R}}\!+\!C_{\mathcal{D}}/\tilde{g}^2}{a_1L_1P}\right)\!{\rm Ei}\!\left(\!-v_{1,i_1}\frac{\sigma^2_{\mathcal{R}}\!+\!C_{\mathcal{D}}/\tilde{g}^2}{a_1L_1P}\right)\!\right]+\sum_{i=1}^{K}\beta^{(K)}_{\alpha_i}\E_{\tilde{g}}\!\left[\exp\!\left(\frac{\sigma^2_{\mathcal{R}}\!+\!C_{\mathcal{D}}/\tilde{g}^2}{L'_ip'_i}\right)\!{\rm Ei}\!\left(-\frac{\sigma^2_{\mathcal{R}}\!+\!C_{\mathcal{D}}/\tilde{g}^2}{L'_ip'_i}\right)\right]\!.\!
	\end{align}
	\hrulefill
	\vspace{-0.25cm}
\end{figure*}

The first expectation term in \eqref{R_ibar}, for the first user, can be evaluated as \eqref{E_f1} shown at the top of the next page, where step $(a)$ is written using \eqref{gamma_f}, and step $(b)$ follows from the cumulative density function (CDF) of the exponential distribution and applying the part-by-part integration equality
\begin{align}\label{pbp}	\setcounter{equation}{25}
&\int_{a}^{b}\log_2(1+cy)f_Y(y)dy=\frac{1}{\ln 2}\Bigg[c\int_{a}^{b}\frac{1-F_Y(y)}{1+cy}dy\nonumber\\
&+(1-F_Y(a))\ln(1+ca)-(1-F_Y(b))\ln(1+cb)\Bigg],
\end{align}
 with $c$ being a constant, and  $f_Y(\cdot)$ and $F_Y(\cdot)$ being the probability density function (PDF) and CDF of the RV $Y$, respectively. 

The first expectation term of the step $(b)$ in \eqref{E_f1}, denoted by $V_1$, can be calculated as
\begin{align}\label{V_1}
&\!\!\!\!\!V_1{=}\E_{\boldsymbol{\mathcal{I}}_1,\tilde{g}}\!\bigg[\!\int_{0}^{\infty}\!\!\!\!\!\!\!{\rm e}^{-y(1\!+\!10^{-s/10})}\!\ln\!\left(\!\!{2y\!\times\!10^{-s\!/\!10}\!\!+\!{\boldsymbol{\mathcal{I}}_1}\!\!+\!\!\frac{C_{\mathcal{D}}}{a_1\!L_1\!P\tilde{g}^2}}\!\right)\!\!dy\!\nonumber\\
&\hspace{0.5cm}-\!\int_{0}^{\infty}\!\!\!\!{\rm e}^{-y(1\!+\!10^{-s/10})}\!\ln\!\left(\!{y\!\times\!10^{-s/10}\!+\!{\boldsymbol{\mathcal{I}}_1}\!+\!\frac{C_{\mathcal{D}}}{a_1\!L_1\!P\tilde{g}^2}}\!\right)\!\!dy\bigg]\nonumber\\
&\!\!\!\!\!\!\!\stackrel{(a)}{=}\!\frac{1}{1\!+\!10^{-s/10}}\!\!\sum_{i_1=1}^{2}\!(-1)^{i_1+1}\E_{\boldsymbol{\mathcal{I}}_1,\tilde{g}}\!\Bigg[\!\exp\!\left(\!v_{1,i_1}\bigg[\!{\boldsymbol{\mathcal{I}}_1}\!+\!\frac{C_{\mathcal{D}}}{a_1\!L_1\!P\tilde{g}^2}\!\bigg]\right)\nonumber\\
&\hspace{3cm}\times\!{\rm Ei}\!\left(\!-v_{1,i_1}\bigg[{\boldsymbol{\mathcal{I}}_1}\!+\!\frac{C_{\mathcal{D}}}{a_1\!L_1\!P\tilde{g}^2}\bigg]\right)\Bigg]\!,\!
\end{align}
where step $(a)$ is obtained using the integral formula \cite[Eq. (4.337.1)]{gradshteyn2014table} with ${\rm Ei}(\cdot)$ being the exponential integral function defined as \cite[Eq. (8.211.1)]{gradshteyn2014table} for negative arguments, and $v_{1,i_1}=(1+10^{s/10})/i_1$, $i_1=1,2$. As it is proven in Appendix C, the expectation term involved in step $(a)$ of \eqref{V_1} can be calculated as \eqref{V1_avgI} shown at the top of the next page where the coefficients for $k=2,3,...,K$ can, recursively, be obtained as
\begin{align}\label{Coeffs}	\setcounter{equation}{28}
\beta^{(k)}_{\alpha_i}&=\frac{-L'_ip'_i\beta^{(k-1)}_{\alpha_i}}{L'_kp'_k-L'_ip'_i},~~~~~i=1,2,...,k-1,\nonumber\\
\beta^{(k)}_{\alpha_k}&=\frac{a_1L_1P\beta^{(k-1)}_{v_{i_1}}}{L'_kp'_kv_{1,i_1}\!-a_1L_1P} +\sum_{i=1}^{k-1}\frac{L'_ip'_i\beta^{(k-1)}_{\alpha_i}}{L'_kp'_k-L'_ip'_i},\nonumber\\
\beta^{(k)}_{v_{i_1}}&=\frac{-a_1L_1P\beta^{(k-1)}_{v_{i_1}}}{L'_kp'_kv_{1,i_1}\!-a_1L_1P}{=\frac{(-a_1L_1P)^k}{\prod_{i=1}^{k}(L'_ip'_iv_{1,i_1}\!-a_1L_1P)}},
\end{align}
with the initial values $\beta^{(1)}_{\alpha_1}=-\beta^{(1)}_{v_{i_1}}=a_1L_1P/(L'_1p'_1v_{1,i_1}\!-a_1L_1P)$. We should remark that this result is obtained with the reasonable assumption $L'_ip'_i\neq L'_jp'_j\neq0$, $\forall i\neq j$, and also $a_1L_1P\neq L'_lp'_lv_{1,i_1}$, $\forall l=1,2,...,K$.

The second expectation term in \eqref{E_f1}, denoted by $V_2$, can be expressed as
\begin{align}\label{V_2}
V_2\stackrel{(a)}{=}&\E_{\boldsymbol{\mathcal{I}}_1,\tilde{g}}\!\Bigg[\!-\exp\!\left(\!{\boldsymbol{\mathcal{I}}_1}\!+\!\frac{C_{\mathcal{D}}}{a_1L_1P\tilde{g}^2}\!\right)\!\!\int_{0}^{\infty}\!\!\!{\rm e}^{-y(1-10^{-s/10})}\nonumber\\
&\hspace{1.4cm}\times\!{\rm Ei}\!\left(\!-2y\!\times\!10^{-s/10}-{\boldsymbol{\mathcal{I}}_1}\!-\!\frac{C_{\mathcal{D}}}{a_1L_1P\tilde{g}^2}\right)\!dy\Bigg]\nonumber\\
&\hspace{-1cm}\stackrel{(b)}{=}\!\frac{1}{1\!-\!10^{-s/10}}\!\!\sum_{i_2=1}^{2}\!(-1)^{i_2+1}\E_{\boldsymbol{\mathcal{I}}_1,\tilde{g}}\!\Bigg[\!\exp\!\left(\!v_{2,i_2}\bigg[\!{\boldsymbol{\mathcal{I}}_1}\!+\!\frac{C_{\mathcal{D}}}{a_1\!L_1\!P\tilde{g}^2}\!\bigg]\right)\nonumber\\
&\hspace{2cm}\times\!{\rm Ei}\!\left(\!-v_{2,i_2}\bigg[{\boldsymbol{\mathcal{I}}_1}\!+\!\frac{C_{\mathcal{D}}}{a_1\!L_1\!P\tilde{g}^2}\bigg]\right)\Bigg]\!,\!
\end{align}
where step $(a)$ follows by calculating the integration over $|\tilde{h}_1|^2$ using \cite[Eq. (3.352.2)]{gradshteyn2014table}, and step $(b)$ is derived using \eqref{Cor1} in Appendix B by defining $v_{2,1}=(1+10^{s/10})/2$ and $v_{2,2}=1$. The expectation term involved in step $(b)$ can be calculated according to \eqref{V1_avgI} and \eqref{Coeffs} by substituting $v_{2,i_2}$ for $v_{1,i_1}$. 
\begin{figure*}
	\normalsize
	\vspace{-0.25cm}
	{
		\begin{align}\label{E_f2}
		&\E_{\boldsymbol{\mathcal{I}}_1,\tilde{g},|\tilde{h}_2|^2,|\tilde{h}_1|^2<|\tilde{h}_2|^2\times10^{-s/10}}\!\!\left[\log_2(1\!+\!\gamma^{(1)}_{\pi_2})\right]\stackrel{(a)}{=}\E_{\boldsymbol{\mathcal{I}}_1,\tilde{g},|\tilde{h}_2|^2}\!\!\left[\int_{0}^{|\tilde{h}_2|^2\times10^{-s/10}}\log_2\!\left(1+\frac{x}{{\boldsymbol{\mathcal{I}}_1}+C_{\mathcal{D}}/(a_1L_1P\tilde{g}^2)}\right)f_{|\tilde{h}_1|^2}(x)dx\right]\nonumber\\
		&\hspace{0.5cm}\stackrel{(b)}{=}\frac{1}{\ln 2}\E_{\boldsymbol{\mathcal{I}}_1,\tilde{g},|\tilde{h}_2|^2}\!\left[\exp\!\left(\!{\boldsymbol{\mathcal{I}}_1}\!+\!\frac{C_{\mathcal{D}}}{a_1L_1P\tilde{g}^2}\!\right)\left\{{\rm Ei}\left(-|\tilde{h}_2|^2\times10^{-s/10}-{\boldsymbol{\mathcal{I}}_1}\!-\!\frac{C_{\mathcal{D}}}{a_1L_1P\tilde{g}^2}\right)-{\rm Ei}\left(-{\boldsymbol{\mathcal{I}}_1}\!-\!\frac{C_{\mathcal{D}}}{a_1L_1P\tilde{g}^2}\right)\right\}\right]\nonumber\\
		&\hspace{1cm}-\frac{1}{\ln 2}\E_{\boldsymbol{\mathcal{I}}_1,\tilde{g},|\tilde{h}_2|^2}\!\left[{\rm e}^{-|\tilde{h}_2|^2\times10^{-s/10}}\ln\!\left(1+\frac{|\tilde{h}_2|^2\times10^{-s/10}}{{\boldsymbol{\mathcal{I}}_1}+C_{\mathcal{D}}/(a_1L_1P\tilde{g}^2)}\right)\right]\nonumber\\
		&\hspace{0.5cm}\stackrel{(c)}{=}\frac{1}{\ln 2}\E_{\boldsymbol{\mathcal{I}}_1,\tilde{g}}\!\left[\exp\!\left(\!{\boldsymbol{\mathcal{I}}_1}\!+\!\frac{C_{\mathcal{D}}}{a_1L_1P\tilde{g}^2}\!\right)\!\times\!\left\{\int_{0}^{\infty}\!\!\!{\rm e}^{-y}{\rm Ei}\!\left(\!-y\!\times\!10^{-s/10}-{\boldsymbol{\mathcal{I}}_1}\!-\!\frac{C_{\mathcal{D}}}{a_1L_1P\tilde{g}^2}\right)\!dy-{\rm Ei}\!\left(\!-{\boldsymbol{\mathcal{I}}_1}\!-\!\frac{C_{\mathcal{D}}}{a_1L_1P\tilde{g}^2}\right)\right\}\right]\nonumber\\
		&\hspace{1cm}+\frac{1}{\ln 2}\!\times\!\frac{1}{1\!+\!10^{-s/10}}\E_{\boldsymbol{\mathcal{I}}_1,\tilde{g}}\!\left[\exp\!\left((1+10^{s/10})\!\times\!\bigg[{\boldsymbol{\mathcal{I}}_1}\!+\!\frac{C_{\mathcal{D}}}{a_1L_1P\tilde{g}^2}\!\bigg]\right){\rm Ei}\!\left(-(1+10^{s/10})\!\times\!\bigg[{\boldsymbol{\mathcal{I}}_1}\!+\!\frac{C_{\mathcal{D}}}{a_1L_1P\tilde{g}^2}\bigg]\right)\right]\nonumber\\
		&\hspace{0.5cm}\stackrel{(d)}{=}\frac{1}{\ln 2}\times\frac{-1}{1+10^{s/10}}\E_{\boldsymbol{\mathcal{I}}_1,\tilde{g}}\!\left[\exp\!\left((1+10^{s/10})\!\times\!\bigg[{\boldsymbol{\mathcal{I}}_1}\!+\!\frac{C_{\mathcal{D}}}{a_1L_1P\tilde{g}^2}\!\bigg]\right){\rm Ei}\!\left(-(1+10^{s/10})\!\times\!\bigg[{\boldsymbol{\mathcal{I}}_1}\!+\!\frac{C_{\mathcal{D}}}{a_1L_1P\tilde{g}^2}\bigg]\right)\right].
		\end{align}
	}
		\hrulefill
	\vspace{-0.25cm}
\end{figure*}

{On the other hand, the second expectation term in \eqref{R_ibar} can be characterized as \eqref{E_f2}, shown at the top of the next page, where step $(a)$ is from the definition of $\gamma^{(1)}_{\pi_2}$ as the dual of \eqref{gamma_s}. Step $(b)$ is obtained by first evaluating the finite integral over $x$ in step $(a)$ using  the part-by-part integration equality in \eqref{pbp} and then applying \cite[Eq. (3.352.1)]{gradshteyn2014table}. Moreover, step $(c)$
	follows from applying \cite[Eq. (4.337.2)]{gradshteyn2014table}  to evaluate the expectation over $|\tilde{h}_2|^2$. Finally, step $(d)$ is written after calculating the integral involved in step $(c)$ using \eqref{Cor1} in Appendix B. Note that ${\rm Ei}(\cdot)$ is a negative quantity for the negative arguments (see, e.g., \cite[Table 1]{harris1957tables}); therefore, the expression obtained in step $(d)$ of \eqref{E_f2} takes positive values.}
Similarly, the expectation term involved in step $(c)$ can be calculated according to \eqref{V1_avgI} and \eqref{Coeffs} by substituting $1+10^{s/10}$ for $v_{1,i_1}$. 
Finally, the expectation over $\tilde{g}$ in \eqref{V1_avgI} can be performed through a one-dimensional integral weighted by $f_{\tilde{g}}(\tilde{g})$ given in \eqref{f_gtild} which, to the best of our knowledge, cannot be calculated in a closed form. 

 The above analysis complete the characterization of the first NOMA user's average individual rate. A similar characterization can be obtained for
  $\mathcal{U}_2$ by substituting $-s$ for $s$ and appropriate change of indexing $1 \leftrightarrow 2$. 

\subsection{Average Sum Rate}
As discussed before, the sum rate of NOMA users
can always be expressed as \eqref{R_sum}, regardless of their decoding order. Therefore, the average sum rate of NOMA users can be calculated as \eqref{R_sum_avg} shown at the top of the next page, where step $(a)$ follows by the definition of instantaneous sum rate in \eqref{R_sum}, applying \cite[Eq. (4.337.2)]{gradshteyn2014table}, and defining ${\rm eEi}(t)\triangleq{\rm e}^t{\rm Ei}(-t)$, $\forall t>0$. Moreover, step $(b)$ is derived by applying  \cite[Eq. (4.337.2)]{gradshteyn2014table} and Eq. \eqref{Cor1} in Appendix B, and then defining $v_{s,1}=10^{s/10}$, $v_{s,2}=1$, $A_{s,1}=(10^{s/10}-1)^{-1}$, and $A_{s,2}=-(1-10^{-s/10})^{-1}$. The expectation term involved in step $(b)$ can be calculated by first applying \eqref{V1_avgI} and \eqref{Coeffs} with substitution of $v_{s,i_s}$ for $v_{1,i_1}$, and then taking a one-dimensional integral over $\tilde{g}$ weighted by $f_{\tilde{g}}(\tilde{g})$.
\begin{figure*}
		\vspace{-0.25cm}
\begin{align}\label{R_sum_avg}
\overline{R}_{\Sigma}\stackrel{(a)}{=}&\E_{\boldsymbol{\mathcal{I}}_1,\tilde{g}}\!\!\left[\int_{0}^{\infty}{\rm e}^{-y}\left\{\log_2\!\left(1+\frac{y\times10^{-s/10}}{{\boldsymbol{\mathcal{I}}_1}+C_{\mathcal{D}}/(a_1L_1P\tilde{g}^2)}\right)-\frac{1}{\ln2}{\rm eEi}\!\left(y\times10^{-s/10}+{\boldsymbol{\mathcal{I}}_1}+C_{\mathcal{D}}/(a_1L_1P\tilde{g}^2)\right)\right\}dy\right]\nonumber\\
\stackrel{(b)}{=}&\frac{1}{\ln2}\sum_{i_s=1}^{2}A_{s,i_s}\E_{\boldsymbol{\mathcal{I}}_1,\tilde{g}}\!\Bigg[\exp\!\left(v_{s,i_s}\bigg[{\boldsymbol{\mathcal{I}}_1}+\frac{C_{\mathcal{D}}}{a_1L_1P\tilde{g}^2}\bigg]\right){\rm Ei}\!\left(-v_{s,i_s}\bigg[{\boldsymbol{\mathcal{I}}_1}+\frac{C_{\mathcal{D}}}{a_1L_1P\tilde{g}^2}\bigg]\right)\Bigg].
\end{align}
\hrulefill
	\vspace{-0.25cm}
\end{figure*}

\subsection{RF-Backhauled System}
In this subsection, we explain how the preceding ergodic capacity analysis can be extended to RF-backhauled systems. As discussed in Section III-C, in the absence of external interference to the destination, the results for the FSO-backhauled system can be extended to that of RF-backhauled one with substituting $C_{\mathcal{D}}^{\rm RF}=N_0/(L_bG_b^2)$ and $\kappa_b=|\tilde{h}_b|^2$ for $C_{\mathcal{D}}$ and $\tilde{g}^2$, respectively. Therefore, the only difference will be the calculation of expectation terms of the form $\E_{\kappa_b}\!\left[\exp\!\left(c'_b[{\sigma^2_{\mathcal{R}}\!+\!C_{\mathcal{D}}^{\rm RF}/\kappa_b}]\right)\!{\rm Ei}\!\left(-c'_b[{\sigma^2_{\mathcal{R}}\!+\!C_{\mathcal{D}}^{\rm RF}/\kappa_b}]\right)\right]$
 instead of $\E_{\tilde{g}}\!\left[\exp\!\left(c'_b[{\sigma^2_{\mathcal{R}}\!+\!C_{\mathcal{D}}/\tilde{g}^2}]\right)\!{\rm Ei}\!\left(-c'_b[{\sigma^2_{\mathcal{R}}\!+\!C_{\mathcal{D}}/\tilde{g}^2}]\right)\right]$, in all of the preceding ergodic capacity analysis, where $\kappa_b$ is distributed according to \eqref{pdf_rice}, and  $c'_b>0$ is a constant, e.g., $c'_b=v_{1,i_1}/(a_1L_1P)$ in the first expectation term of \eqref{V1_avgI}. As we know, such expectations cannot be characterized in closed forms and need to be calculated through  one-dimensional integrals weighted by $f_{\kappa_b}(\kappa_b)$. 
 
 Moreover, when the destination is subject to the presence of $K_{\rm d}$ interfering users, $C_{\mathcal{D}}/\tilde{g}^2$ in all of the previously-derived ergodic capacity formulas should be replaced by $(\boldsymbol{\mathcal{I}}''+N_0)/(L_bG_b^2\kappa_b)=C_{\mathcal{D}}^{\rm RF}/\kappa_b+\boldsymbol{\mathcal{I}}''/(L_bG_b^2\kappa_b)$, where $\boldsymbol{\mathcal{I}}''\triangleq\sum_{j=1}^{K_{\rm d}}{L''_jp''_j}|\tilde{h}''_j|^2$ is the total external interference to the destination. 
 Therefore, in order to characterize the ergodic capacity, instead of calculating $\E_{\tilde{g}}\!\left[\exp\!\left(c'_b[{\sigma^2_{\mathcal{R}}\!+\!C_{\mathcal{D}}/\tilde{g}^2}]\right)\!{\rm Ei}\!\left(-c'_b[{\sigma^2_{\mathcal{R}}\!+\!C_{\mathcal{D}}/\tilde{g}^2}]\right)\right]$, we have to evaluate
 \begin{align}\label{Eb}
 {E}_{b}\triangleq\E_{\kappa_b,\boldsymbol{\mathcal{I}}''}\!\Big[&\!\exp\!\left(c'_b\left[{\sigma^2_{\mathcal{R}}\!+C_{\mathcal{D}}^{\rm RF}/\kappa_b\!+\!\boldsymbol{\mathcal{I}}''/(L_bG_b^2\kappa_b)}\right]\right)\nonumber\\
 &\hspace{-1cm}\times{\rm Ei}\!\left(-c'_b\left[{\sigma^2_{\mathcal{R}}\!+C_{\mathcal{D}}^{\rm RF}/\kappa_b\!+\!\boldsymbol{\mathcal{I}}''/(L_bG_b^2\kappa_b)}\right]\right)\!\Big].
 \end{align}
By defining $B_{\kappa_b}\triangleq\sigma^2_{\mathcal{R}}\!+C_{\mathcal{D}}^{\rm RF}/\kappa_b$ and $\alpha'_j(\kappa_b)\triangleq L''_jp''_j/(L_bG_b^2\kappa_b)$, $j=1,2,...,K_{\rm d}$, and then applying a similar approach to Appendix C, one can show that $E_b$ in \eqref{Eb} can be calculated as \eqref{Eb_avg1} shown at the top of the next page
\begin{figure*}[!t]
	\normalsize
		\vspace{-0.25cm}
	\begin{align}\label{Eb_avg1}
\!\!E_b\!=\!\E_{\kappa_b}\!\!\left[\beta^{(K_{\rm d})}_{c'_b}\exp\!\left(\!c'_b\!\left[\sigma^2_{\mathcal{R}}\!+\!\frac{C_{\mathcal{D}}^{\rm RF}}{\kappa_b}\right]\!\right)\!{\rm Ei}\!\left(\!\!-c'_b\!\left[\sigma^2_{\mathcal{R}}\!+\!\frac{C_{\mathcal{D}}^{\rm RF}}{\kappa_b}\right]\!\right)\!\right]\!+\!\sum_{j=1}^{K_{\rm d}}\E_{\kappa_b}\!\!\left[\beta^{(K_{\rm d})}_{\alpha'_j}\exp\!\left(\!\frac{N_0\!+\!\sigma^2_{\mathcal{R}}L_bG_b^2\kappa_b}{L''_jp''_j}\!\right)\!{\rm Ei}\!\left(\!-\frac{N_0\!+\!\sigma^2_{\mathcal{R}}L_bG_b^2\kappa_b}{L''_jp''_j}\!\right)\!\right]\!.\!
	\end{align}
	\hrulefill
		\vspace{-0.25cm}
\end{figure*}
where the coefficients for $k=2,3,...,K_{\rm d}$ are, recursively, defined as
\begin{align}\label{Coeffs2}
\beta^{(k)}_{\alpha'_j}&=\frac{-L''_jp''_j\beta^{(k-1)}_{\alpha'_j}}{L''_kp''_k-L''_jp''_j},~~~~~j=1,2,...,k-1,\nonumber\\
\beta^{(k)}_{\alpha'_k}&=\frac{L_bG_b^2\kappa_b\beta^{(k-1)}_{{c'_b}}}{L''_kp''_kc'_b-L_bG_b^2\kappa_b} +\sum_{j=1}^{k-1}\frac{L''_jp''_j\beta^{(k-1)}_{\alpha'_j}}{L''_kp''_k-L''_jp''_j},\nonumber\\
\beta^{(k)}_{c'_b}&=\frac{-L_bG_b^2\kappa_b\beta^{(k-1)}_{c'_b}}{L''_kp''_kc'_b-L_bG_b^2\kappa_b}{=\frac{(-L_bG_b^2\kappa_b)^k}{\prod_{i=1}^{k}(L''_ip''_ic'_b-L_bG_b^2\kappa_b)}},
\end{align}
with the initial values $\beta^{(1)}_{\alpha'_1}=-\beta^{(1)}_{c'_b}=L_bG_b^2\kappa_b/(L''_1p''_1c'_b-L_bG_b^2\kappa_b)$. Finally, the ergodic capacity formulas for the RF-backhauled system can be obtained after one-dimensional integrals over $\kappa_b$ to perform the expectations of the form expressed in \eqref{Eb_avg1}. We should again emphasize that this result is valid under the reasonable assumption $L''_ip''_i\neq L''_jp''_j\neq0$, $\forall i\neq j$, and also $L_bG_b^2\kappa_b\neq L''_lp''_lc'_b$, $\forall l=1,2,...,K_{\rm d}$.

{\textbf{Remark 4:} Our characterization of the average individual- and sum-rate formulas for both FSO- and RF-backhauled systems are up to only one-dimensional integrals over the fading coefficient of the backhaul link. That is equivalent to say that the ergodic capacity closed-form expressions are obtained for the single-hop uplink NOMA subject to some exterior multiuser interference, or better to say, for the considered dual-hop system model given each realization of the backhaul fading coefficient.}

\section{Numerical Results}
In this section, we present the numerical results to evaluate the performance of uplink NOMA over mixed RF-FSO and dual-hop RF/RF systems, and corroborate the correctness of the derived outage probability and ergodic capacity formulas. Some of the  parameters considered for simulations (unless explicitly specified) are listed in Table I. For the multiuser interference, we consider the product of $L'_kp'_k$, $k=1,2,...,K$, to be the $k$-th element of the vector $K_{\mathcal{I},R}P_0L_2{\boldsymbol{u}'_{10}}$ where $P_0=1$ \si{mW}, $K_{\mathcal{I,R}}\geq0$ is a constant to define the upper bound of the received power from each interfering user to the relay as a factor of $P_0L_2$, and ${\boldsymbol{u}'_{10}}=(0.6957,\!0.6279,\!0.4504,\!0.4736,\!0.9497, \!0.0835,0.2798,0.4470,\\0.5876,0.8776)$ is a length-10 vector of uniformly generated numbers over the interval $(0,1)$. Similarly, we consider the external interference to the RF receiver of the destination in the case of the RF-backhauled system to be of the form $K_{\mathcal{I},D}P_0L_2{\boldsymbol{u}''_{10}}$ with the constant $K_{\mathcal{I},D}\geq0$ {defining the upper bound of the received power from each interfering user to the destination as a factor of $P_0L_2$ and ${\boldsymbol{u}''_{10}}=(0.5259,\!0.9635,\!0.5688,\!0.2584,\!0.2959,\!0.7439,0.9797,0.3491,\\0.8371,0.5587)$ being another length-10 vector of uniformly generated numbers over the interval $(0,1)$.}
\begin{table}[t]
	\centering
	\caption{Some of the important parameters used for simulations {\cite{jamali2016link}}.}
	\label{T2}
	\begin{tabular}{M{2.33in}||M{0.5in}}
		
		Coefficient & Value\\
		\hline \hline
		Responsivity of the photodetector, $\rho$ & $0.5$ \si{V^{-1}}\\\hline
		Electrical-to-optical conversion coefficient, $\eta$ & $1$\\\hline
		Optical receiver aperture radius, $r$ & $10$ \si{cm}\\\hline
		Laser beam divergence angle, $\phi$ & $2$ \si{mrad}\\\hline
		Noise power at the relay and destination RF receivers, $\sigma^2_{\mathcal{R}}$ and $N_0$ &$-80$ \si{dBm} \\\hline
		Noise variance at the destination FSO receiver, $\sigma^2_{\mathcal{D}}$ & $10^{-14}$ \si{A^2}\\\hline
		Reference distance of the RF link, $d^{\mathrm{RF}}_{\mathrm{ref}}$ & $80$ \si{m}\\\hline
		Number of interfering users to the relay, $K$ & $10$\\
		\hline
		Number of interfering users to the destination, $K_{\rm d}$ & $10$\\
		\hline
		Transmitter and receiver antenna gains of the user-relay RF links, $(G^{\mathrm{RF}}_{t,i},G^{\mathrm{RF}}_{r,\mathcal{R}})$  & $(5,8)$ \si{dBi}\\\hline
		Transmitter and receiver antenna gains of the relay-destination RF backhaul link, $(G^{\mathrm{RF}}_{t,\mathcal{R}},G^{\mathrm{RF}}_{r,\mathcal{D}})$  & $(10,15)$ \si{dBi}\\\hline
		Number of iterations for numerical simulations, $N_t$ & $5\times10^{6}$\\\hline
		Length of the FSO backhaul link, $d_{\mathcal{RD}}$& $1200$ \si{m}\\\hline
 Wavelength of FSO signal, $\lambda^{\mathrm{FSO}}$ & $1550$ \si{nm}\\\hline
 Frequency of RF signal, $f^{\rm RF}$ & $3$ \si{GHz}\\\hline
 Parameter of the Rician distribution for the RF backhaul link, $\Omega$ & $6$ \si{dB}\\\hline
 Path-loss exponent of RF links, $\nu$ & $3.5$
	\end{tabular}
	  \vspace{-0.2in}
\end{table}

\begin{figure}\label{ff1}
	\centering
	\includegraphics[width=3.6in]{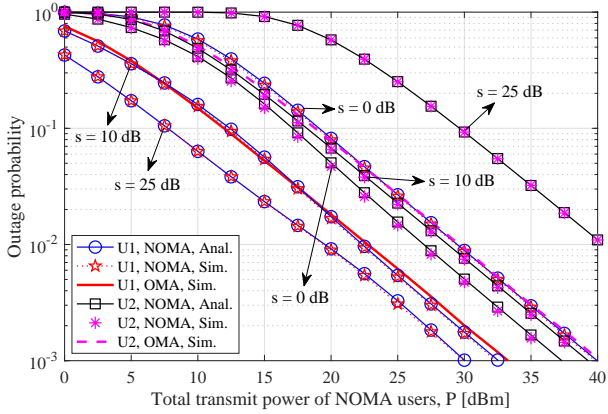}
	\caption{Individual-rate outage probability results of the mixed RF-FSO NOMA system for three values of the power back-off step $s=0$, $10$, and $25$ \si{dB}. The other specific parameters are $\gamma_{\rm th}^{(1)}=0.8$ ($R_{\rm th}^{(1)}=0.8480$), $\gamma_{\rm th}^{(2)}=0.4$ ($R_{\rm th}^{(2)}=0.4854$), $d^{\mathrm{RF}}_1=100$ \si{m}, $d^{\mathrm{RF}}_2=200$ \si{m}, $K_{\mathcal{I,R}}=1$, $\kappa=0.43\times10^{-3}$ $\rm{m^{-1}}$ (clear air), $\alpha=4$, $\beta=2$, $\zeta=2$, and $G=100$.}
	\vspace{-0.15in}
\end{figure}
Fig. 2 shows the individual-rate outage performance of the uplink mixed RF-FSO NOMA system for three different values of the power back-off step $s=0$, $10$, and $25$ \si{dB}. We assume that the first and second users are in distances of $d^{\mathrm{RF}}_1=100$ \si{m} and $d^{\mathrm{RF}}_2=200$ \si{m} from the relay, respectively; this together with the values in Table I implies $L_1=9.04\times10^{-8}$ and $L_2=8\times10^{-9}$. The other adopted parameters are listed in the caption of Fig. 2.
For $s=0$ we will have $a_1L_1=a_2L_2$; therefore, one should expect lower outage probabilities for the second NOMA user given its lower threshold SINR $\gamma_{\rm th}^{(2)}=0.4$ (equivalently, lower target rate $R_{\rm th}^{(2)}=0.485$ \si{bits/sec/Hz}). However, by increasing $s$ a larger fraction of power will be assigned to the first NOMA user, and $\mathcal{U}_1$ achieves lower outage probabilities even if it demands a larger SINR threshold. As a consequence, increasing $s$ will decrease the outage probability of $\mathcal{U}_1$ and increase the outage probability of $\mathcal{U}_2$. Moreover, the match between the analytical results and Monte-Carlo simulations corroborates the correctness of the derived closed-form expressions for the individual-rate outage probabilities. 

The comparison between NOMA and orthogonal multiple access (OMA) is also depicted in Fig. 2. We assume, for the OMA transmission, that the total transmission time is equally divided between the two users and each user employs the entire transmission power $P$ during its corresponding time slot. Therefore, denoting the SINR of the $i$-th OMA user
 by $\gamma^{(i)}_{\rm OMA}$, $i\in\{1,2\}$, each OMA user will have the rate $R^{(i)}_{\rm OMA}=0.5\log_2(1+\gamma^{(i)}_{\rm OMA})$.
 Then it is easy to verify that, in order to achieve the target data rate $R_{\rm th}^{(i)}=\log_2(1+\gamma_{\rm th}^{(i)})$, each $i$-th OMA user has to satisfy the threshold SINR of $\gamma^{(i)}_{{\rm th},\rm OMA}=(1+\gamma_{\rm th}^{(i)})^2-1$. It is observed that NOMA transmission is in favor of the first user except for very small values of $s$ while the second user experiences an opposite situation. In particular for $s=10$ \si{dB} where $a_1L_1=10a_2L_2$, i.e., when the roughly an order of magnitude difference between the channel gains $L_1$ and $L_2$ is reflected in the NOMA power allocation policy, both of the NOMA users (especially the second user) achieve better performance than the OMA transmission. This suggests the possible existence of some optimal values for $s$ (see also Fig. 5) given some specific figures of merit. 

\begin{figure}\label{fig2}
	\centering
	\includegraphics[width=3.6in]{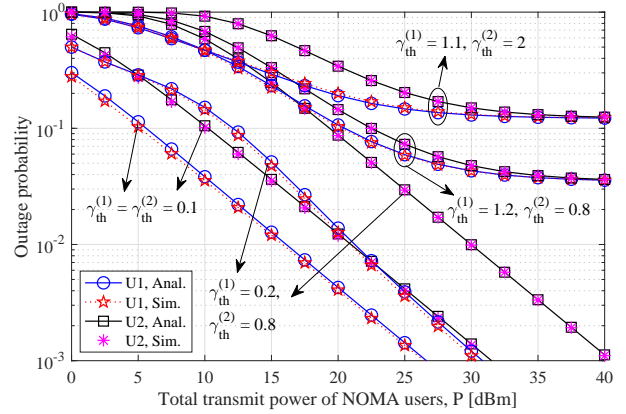}
	\caption{Individual-rate outage probability results of the mixed RF-FSO NOMA system for the power back-off step  $s=5$ \si{dB}, and different values of the threshold SINRs. The other parameters are  the same as Fig. 2.}
	\vspace{-0.15in}
\end{figure}
Fig. 3 illustrates the individual-rate outage performance of the mixed RF-FSO NOMA system for $s=5$ \si{dB} and different values of the threshold SINRs. As expected, outage performance degrades with increasing the threshold SINRs. More importantly, the induced interference between NOMA users due to the non-orthogonal operation limits the outage performance for the large values of the threshold SINRs and prevents achieving small enough outage probabilities even for the large values of the transmitted power. As a result,  the system outage performance saturates while the saturation limits are higher for the larger values of the threshold SINRs. {Note that the channel is stochastic and the decoding order of the NOMA users changes dynamically. Therefore, the first-decoded user dynamically changes between $\mathcal{U}_1$ and $\mathcal{U}_2$; hence, both NOMA users experience interference  and cannot meet large values of the threshold SINRs even if $P$ is increased.}

\begin{figure}\label{fig4}
	\centering
	\includegraphics[width=3.6in]{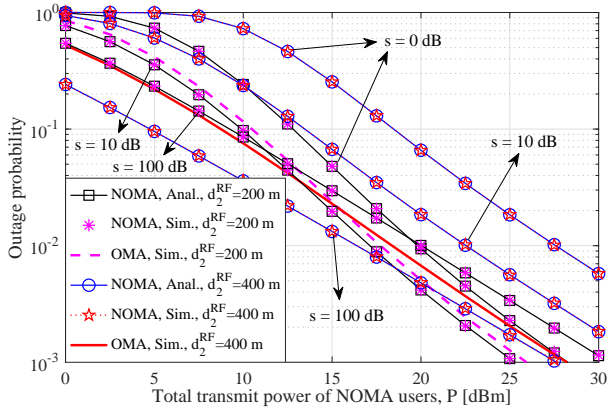}
	\caption{Sum-rate outage probability results of the mixed RF-FSO NOMA system for $\gamma^{\Sigma}_{\rm th}=1.2$, three different power back-off steps  $s=0$, $10$, and $100$ \si{dB}, and two different distances of the second user $d^{\mathrm{RF}}_2=200$ and $400$ \si{m}. The other parameters are  the same as Fig. 2.}
	\vspace{-0.15in}
\end{figure}
Sum-rate outage probability results of the mixed RF-FSO NOMA system are depicted in Fig. 4 for $\gamma^{\Sigma}_{\rm th}=1.2$, three different power back-off steps  $s=0$, $10$, and $100$ \si{dB}, and two different distances of the second user $d^{\mathrm{RF}}_2=200$ and $400$ \si{m}. The other parameters are  the same as Fig. 2. For the sake of comparison, the simulation results are also provided for the OMA sum-rate outage probabilities. Similar to Fig. 2, we assume that the transmission time is equally divided between the two users while each user transmits with the full power $P$ during its transmission time. Then it is easy to verify that the sum-rate outage probability of the OMA transmission in satisfying the target rate $R^{\Sigma}_{\rm th}=\log_2(1+\gamma^{\Sigma}_{\rm th})$ can be obtained as
\begin{align}\label{oma}
P^{\Sigma}_{\rm out,OMA}\!\!=\!\Pr\!\left(\!\!\sqrt{\!(1\!+\!\gamma^{(1)}_{\rm OMA})(1\!+\!\gamma^{(2)}_{\rm OMA})}\!-\!1\!<\!\gamma^{\Sigma}_{\rm th}\right).
\end{align}
For the case of $d^{\mathrm{RF}}_2=2d^{\mathrm{RF}}_1=200$ \si{m}, which corresponds to $L_1=9.04\times10^{-8}$ and $L_2=8\times10^{-9}$, one can see that by increasing $s$ the sum-rate outage probability decreases first and then increases, at least for higher transmit powers. Specifically, for $s=10$ \si{dB} where $a_1L_1=10a_2L_2$, i.e., when the roughly an order of magnitude difference between the channel gains $L_1$ and $L_2$ is reflected in the NOMA power allocation policy, NOMA outperforms OMA in terms of the sum-rate outage probability. One can easily verify that in the special case of second user having a very poor channel condition compared to the first user, NOMA, with assigning the whole power to the first user (equivalently, $s\to\infty$), can achieve twice the OMA rate. This can be observed also from the case of $d^{\mathrm{RF}}_2=4d^{\mathrm{RF}}_1=400$ \si{m}, corresponding to $L_1=9.04\times10^{-8}$ and $L_2=7.06\times10^{-10}$, with the choice of $s=100$ \si{dB} (at least at reasonably low transmit powers where $R^{(2)}_{\rm OMA}=0.5\log_2(1+\gamma^{(2)}_{\rm OMA})\approx0$).

\begin{figure}\label{fig5}
	\centering
	\includegraphics[width=3.6in]{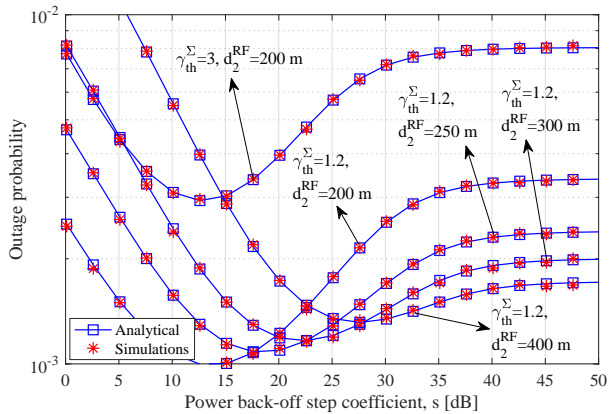}
	\caption{Sum-rate outage probability results of the mixed RF-FSO NOMA system for $P=25$ \si{dBm}, and different values of $d^{\mathrm{RF}}_2$ and $\gamma^{\Sigma}_{\rm th}$. The other parameters are  the same as Fig. 2.}
	\vspace{-0.15in}
\end{figure}
To better observe the trend of sum-rate outage probability as a function of $s$, we have characterized in Fig. 5 the sum-rate outage performance of the system for $P=25$ \si{dBm}, and different values of $d^{\mathrm{RF}}_2$ and $\gamma^{\Sigma}_{\rm th}$.
It is observed that the outage performance decreases first and then increases with the increase of $s$. 
Therefore, given any set of system parameters, there is a power back-off step $s^*$ minimizing the sum-rate outage probability. However, such a $s^*$ is not necessarily the best operation point as such an operation region may depend to the individual outage probabilities and achievable rates and not only to the sum-rate outage probability. By comparing the four plots corresponding to $\gamma_{\rm th}^{\Sigma}=1.2$ we can observe that increasing $d^{\mathrm{RF}}_2$, i.e., differentiating more between the channel qualities of the NOMA users, shifts $s^*$ to the larger values. Moreover, the outage probability is higher for the larger values of $d^{\mathrm{RF}}_2$ before the minimum of $s^*$'s and is lower after their maximum. Furthermore, we have observed that increasing $\gamma_{\rm th}^{\Sigma}$, while $d^{\mathrm{RF}}_2$ is kept fixed, increases the outage probability and does not change the outage-minimizing $s$. 
{ Finally, it can be observed that as $s$ increases, we have $a_1\to 1$ and $a_2\to 0$. Therefore, further increasing $s$, beyond a certain large threshold, does not noticeably change the power allocation coefficients. Consequently, the outage probability saturates while increasing $s$. }

\begin{figure}\label{fig7}
	\centering
	\includegraphics[width=3.6in]{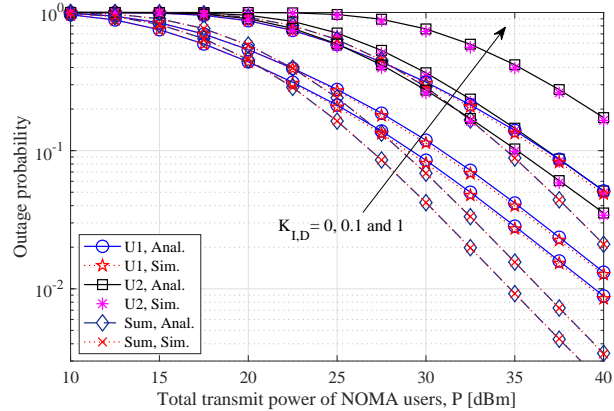}
	\caption{Individual- and sum-rate outage probability results of the dual-hop RF/RF NOMA system for $d^{\mathrm{RF}}_1=100$ \si{m}, $d^{\mathrm{RF}}_2=200$ \si{m}, $d_{\mathcal{RD}}=500$ \si{m}, $\gamma^{(1)}_{\rm th}=0.8$, $\gamma^{(2)}_{\rm th}=0.4$, $\gamma^{\Sigma}_{\rm th}=1.2$, $s=10$ \si{dB}, $G_b=1000$, $K_{\mathcal{I,R}}=1$, and three different values of $K_{\mathcal{I,D}}=0$, $0.1$ and $1$.}
	\vspace{-0.15in}
\end{figure}
The individual- and sum-rate outage probability results of the dual-hop RF/RF NOMA system {with $d_{\mathcal{RD}}=500$ \si{m}}  are characterized in Fig. 6 for three different values of $K_{\mathcal{I,D}}=0$, $0.1$ and $1$. As observed, even small values of $K_{\mathcal{I,D}}$ result in significant performance degradation, mainly due to the weak desired signal at the destination.
 This figure, in addition to confirming the accuracy of the results derived in Section III-C,  highlights the superiority of FSO backhauling in terms of outage performance (compare the case of $K_{\mathcal{I,D}}=0$ in Fig. 6 with the corresponding plots in Figs. 2 and 4, and note that the length of backhaul link here is much less than that of FSO backhaul), especially for longer backhaul ranges, noticing that FSO links with even lower SINRs may be preferred given their much more available bandwidths compared to RF backhaul links. { We should emphasize that increasing the number of interfering users increases the outage probability in a similar fashion. However, the results are not included in this paper due to the space limitation.}

\begin{figure}\label{fig8}
	\centering
	\includegraphics[width=3.6in]{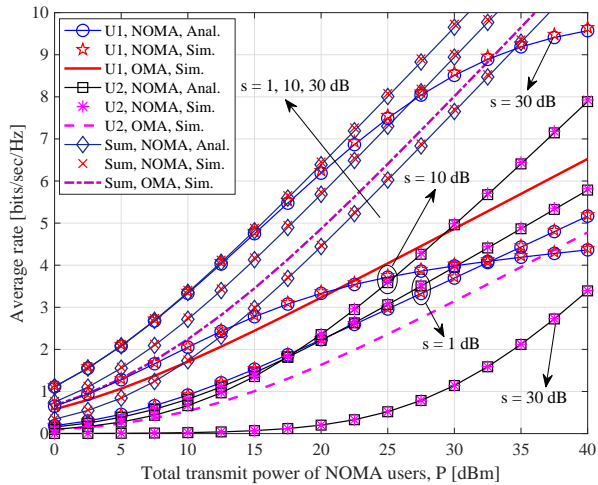}
	\caption{Average individual and sum rates of the mixed RF-FSO NOMA system for three values of the power back-off step $s=0$, $10$, and $30$ \si{dB}. The other parameters are the same as Fig. 2.}
	\vspace{-0.15in}
\end{figure}
Average individual and sum rates of the mixed RF-FSO NOMA system are  characterized in Fig. 7 for $d^{\mathrm{RF}}_1=100$ \si{m}, $d^{\mathrm{RF}}_2=200$ \si{m}, $d_{\mathcal{RD}}=1200$ \si{m}, and three values of the power back-off step $s=0$, $10$, and $30$ \si{dB}.
 It is observed that, at relatively low transmit powers, increasing $s$ (within the range of $s$ in this figure) increases the rate of the first NOMA user and decreases the rate of the second user while the average sum rate is also increased. This is because $a_1L_1=a_2L_2\times10^{s/10}$ and $s\geq0$; therefore, more power is assigned to the first NOMA user for the larger values of $s$ which increases the first user's rate. On the other hand, for the larger values of $s$ with higher probabilities the decoding order is $\pi_1$, i.e., the first user is decoded first in the presence of interference signal from the second NOMA user; hence, by increasing $P$ the interference power to the first user increases limiting its performance and saturating its average rate at larger $P$'s. This is while the second NOMA user is usually decoded after removing the interference signal of the first user, especially for larger values of $s$; hence, the average rate of the second NOMA user can increase without saturation. 
Given the increasing trend of the rate of the second NOMA user and the saturating trend of the first NOMA user, the average rate curves of these two users intersect at some power $P$ such that the first user has a higher rate before that crossing point and the second user outperforms after that power. Also as $s$ increases, this crossing point shifts to the right meaning that the first NOMA user has higher data rates over wider ranges of the transmit power for the larger values of $s$.
We have also included the simulation results of the OMA average rates. The simulation results confirm the superiority of NOMA over OMA in terms of individual and sum rates (see, e.g., the results for $s=10$ \si{dB}). 

\begin{figure}\label{fig9}
	\centering
	\includegraphics[width=3.6in]{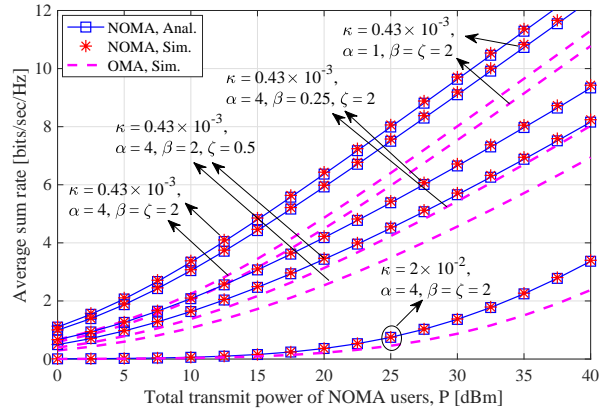}
	\caption{Average sum rate of the mixed RF-FSO system for $s=25$ \si{dB}, and different values of FSO link parameters $\kappa$, $\alpha$, $\beta$, and $\zeta$. The other parameters are the same as Fig. 2.}
	\vspace{-0.15in}
\end{figure}
The impact of the FSO link parameters on the average sum rate of the NOMA and OMA mixed RF-FSO systems is investigated in Fig. 8 for $s=25$, $d^{\mathrm{RF}}_1=100$ \si{m}, $d^{\mathrm{RF}}_2=200$ \si{m}, $d_{\mathcal{RD}}=1200$ \si{m}, and different values of $\kappa$, $\alpha$, $\beta$, and $\zeta$. It is observed that the average sum rate decreases by increasing $\kappa$ (equivalent to having larger path-losses) and decreasing $\alpha$, $\beta$, and $\zeta$ (equivalent to stronger atmospheric turbulence). As a result, in the extreme atmospheric conditions the performance of FSO backhauling might be inferior to that of RF backhauling; this necessitates adaptive switching between RF and FSO infrastructure to simultaneously take the advantage of the potentials of both FSO (e.g., higher data rates and longer ranges) and RF (e.g., resilience to atmospheric adverse conditions) backhauling. Furthermore, note that for all of the considered scenarios NOMA average sum rate outperforms that of the OMA transmission. 

\begin{figure}\label{fig11}
	\centering
	\includegraphics[width=3.6in]{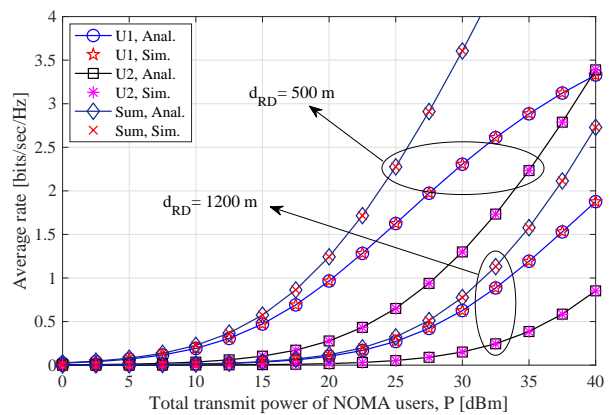}
	\caption{Average rates of the dual-hop RF/RF system for $s=8$ \si{dB},  and two different values of $d_{\mathcal{RD}}=500$ and $1200$ \si{m}. The other parameters are the same as Fig. 6.}
	\vspace{-0.15in}
\end{figure}
Average individual and sum rates of the dual-hop RF/RF system are characterized in Fig. 9 for $s=8$ \si{dB}, $d^{\mathrm{RF}}_1=100$ \si{m}, $d^{\mathrm{RF}}_2=200$ \si{m}, and two different values of $d_{\mathcal{RD}}=500$ and $1200$ \si{m}. It is observed that as the backhaul link length increases the point for which the individual rates intersect (see, e.g., Fig. 7) moves to the right hand side, i.e., the first NOMA user will have higher average rates over a wider range of the transmit powers. This figure, in addition to confirming the accuracy of the results derived in Section IV-C, further highlights the superiority of FSO backhauling in terms of average achievable rates, especially for longer backhaul ranges, given much more available bandwidth of optical links. 

\section{Conclusions and Future Directions}
We studied the performance of uplink NOMA over mixed RF-FSO systems where the NOMA users employ RF signals for concurrent transmission to an intermediate relay which forwards the amplified version of the received signal in the presence of multiuser interference to the destination through an ultra high-speed FSO link. We adopted dynamic-order decoding to determine the priority of the users at the destination based on their instantaneous CSI. The analysis of the paper carried out in a general model of the channel and system, such as including different aspects of the RF and FSO channels and taking into account the presence of multiuser interference, e.g., due the co-channel interference from some nearby users aiming to communicate to some other relays or destinations. We also extended the results to the case of RF-backhauled systems, i.e., the conventional dual-hop RF/RF systems in the presence of multiuser interference. 
In terms of the outage probability performance, we derived the closed-form expressions for both mixed RF-FSO and dual-hop RF/RF NOMA systems.
Moreover, we characterized the average individual and sum rates of both FSO- and RF-backhauled systems up to only one-dimensional integrals over the fading coefficient of the backhaul link. We presented extensive numerical results confirming the accuracy of the derived outage probability and average rate formulas and characterizing the impact of various design and channel parameters on the system performance.

Our results revealed the superiority of FSO backhauling compared to RF backhauling in terms of outage probability and ergodic capacity. This advantage can further be highlighted, as a future work, by analyzing the throughput of the dual-hop RF-FSO and RF/RF NOMA systems (e.g., when the users of the access links have enough data for transmission and the backhaul link is imposing a bottleneck) by taking into account the much higher available bandwidth of optical links. Moreover, given the sensitivity of FSO links to the atmospheric conditions (see, e.g., Fig. 8), it is inevitable to design hybrid RF/FSO backhaul links to simultaneously take the advantages of both FSO and RF systems; this necessitates further studies to characterize the performance of relay-assisted  NOMA systems when the backhaul link dynamically switches between FSO and RF systems given properly-defined switching policies. More importantly, the results of this paper provide novel expressions for the outage probabilities and ergodic capacities of dual-hop RF-FSO and RF/RF systems subject to some external interference to the RF receivers. This is particularly relevant to mmWave NOMA systems which recently have received extensive attention. In fact, in mmWave NOMA systems, inter-beam interference due to the side-lobes of nearby mmWave beams can adversely affect the power-domain NOMA users grouped over a given mmWave beam. As a result, the analysis in this paper can significantly pave the way toward future research on mmWave NOMA systems.
 
\appendices
{\section{Closed-Form Expression for Eq. \eqref{E_I''}}
In order to derive the closed form of Eq. \eqref{E_I''}, we need to calculate integrals of the form
\begin{align}\label{J1,1}
\mathcal{J}_1\triangleq \int_{0}^{\infty}\frac{x^n\times{\rm e}^{-ax-b/x}}{x+c}I_0\!\left(\sqrt{dx}\right)dx,
\end{align}
for some positive constants $a,b,c,d$, and integer $n$.\footnote{The use of symbols in this appendix should not be confused with the rest of the paper as we only reuse some symbols during the proof steps.} Using \cite[Eq. (8.447.1)]{gradshteyn2014table} to expand $I_0\!\left(\sqrt{dx}\right)$ as an infinite series, we can rewrite \eqref{J1,1} as
\begin{align}\label{J1,2}
\mathcal{J}_1=\sum_{k=0}^{\infty}\frac{d^k}{4^k(k!)^2}\underbrace{\int_{0}^{\infty}\frac{x^{n+k}\times{\rm e}^{-ax-b/x}}{x+c}dx}_{\mathcal{J}_2}.
\end{align}
Hence, the problem boils down to the closed-form calculation of $\mathcal{J}_2$ defined in \eqref{J1,2}. To do so, we first use \cite[Eq. (07.34.03.0228.01)]{wolfrm} together with \cite[Eq. (9.301)]{gradshteyn2014table} to write $\exp(-b/x)$ in the integral form
\begin{align}\label{exp}
\exp(-b/x)=\frac{1}{2\pi j}\oint\limits_{\mathcal{C}}\Gamma(-s)\left(\frac{x}{b}\right)^{-s}ds,
\end{align}
where $\mathcal{C}$ is a complex contour of integration ensuring the convergence of the above Mellin-Barnes type integral, e.g., a vertical line starting from the point $-\epsilon-j\infty$ and terminating at the point $-\epsilon+j\infty$ for any $\epsilon>0$.\footnote{Based on \cite[Sec. 1.1]{kilbas2004h}, this is a proper choice for $\mathcal{C}$ that separates all the poles (i.e., all nonnegative integers) to the right, as required for the convergence of \eqref{exp}.} Now, by substituting \eqref{exp} in $\mathcal{J}_2$ we have
\begin{align}\label{J2,2}
\mathcal{J}_2&=\frac{1}{2\pi j}\oint\limits_{\mathcal{C}}\frac{\Gamma(-s)}{b^{-s}}\left\{\int_{0}^{\infty}\frac{x^{n+k-s}\times{\rm e}^{-ax}}{x+c}dx\right\}ds\nonumber\\
&{\hspace{-0.4cm}}\stackrel{(a)}{=}\!\frac{c^{n+k}{\rm e}^{ac}}{2\pi j}\!\!\oint\limits_{\mathcal{C}}\!\left(\frac{c}{b}\right)^{\!\!-s}\!\Gamma(-s)\Gamma(n\!+\!k\!+\!1\!-\!s)\Gamma(s\!-\!n\!-\!k,\!ac)ds\nonumber\\
&\stackrel{(b)}{=}c^{n+k}{\rm e}^{ac}\mathcal{H}_{2,1}^{1,2}\left[\frac{c}{b}{\bigg |\begin{matrix}
	(1,1,0),(-n-k,1,0)\\ (-n-k,1,ac)
	\end{matrix}}\right]
\nonumber\\
&\stackrel{(c)}{=}c^{n+k}{\rm e}^{ac}\mathcal{H}_{1,2}^{2,1}\left[\frac{b}{c}{\bigg |\begin{matrix}
	(n+k+1,1,ac)\\(0,1,0),(n+k+1,1,0)
	\end{matrix}}\right]
\nonumber\\
&\stackrel{(d)}{=}c^{n+k}{\rm e}^{ac}\mathsf{G}_{1,2}^{2,1}\left[\frac{b}{c}{\bigg |\begin{matrix}
	(n+k+1,ac)\\(0,0),(n+k+1,0)
	\end{matrix}}\right],
\end{align}
where $(a)$ is calculated using \cite[Eq.
(3.383.10)]{gradshteyn2014table} when noting that $\operatorname{Re}(n+k+1-s)=n+k+\epsilon+1>0$, and $(b)$ is obtained using the definition of the generalized upper incomplete Fox's H-function $\mathcal{H}_{.,.}^{.,.}$ \cite{yilmaz2009product}. Note that the upper incomplete gamma function $\Gamma(s\!-\!n\!-\!k,\!ac)$ does not have any poles, and the poles of the other two gamma functions in $(a)$, i.e., $\Gamma(-s)$ and $\Gamma(n+k+1-s)$ are at $l$ and $k+n+1+l$, respectively, for any nonnegative integer $l$. Therefore, all the poles of $\Gamma(-s)$ and $\Gamma(n+k+1-s)$ lie on the right side of the vertical line considered above for $\mathcal{C}$; thus, according to \cite{yilmaz2009product}, the aforementioned vertical line for the contour $\mathcal{C}$ also guarantees the convergence of the Mellin-Barnes type integral in $(a)$. We should further mention that step $(c)$ is obtained using \cite[Eq. (A.9)]{yilmaz2009product}, and step $(d)$ is from the definition of the upper incomplete Meijer's G-function $\mathsf{G}_{.,.}^{.,.}$, which is a special case of $\mathcal{H}_{.,.}^{.,.}$ obtained when the second element of all triples corresponding to the coefficients of $s$ or $-s$ in the arguments of (incomplete) gamma functions are equal to one. Finally, by inserting $n=K_{\rm d}$, $a=1+\Omega$, $b=BC_{\mathcal{D}}^{\rm RF}$, $c=BL''_lp''_l/(L_bG_b^2)$, and $d=4\Omega(1\!+\!\Omega)$ in \eqref{J2,2} and then \eqref{J1,2}, we get the closed-form expression in \eqref{E_I''}.
}
\section{Useful Integral Equations over ${\rm Ei}(\cdot)$}
{Lemma 1, Corollary 1, and Corollary 2 can be found in the literature with slight variations. However, independent proofs are included for the sake of completeness.}

\textbf{Lemma 1:} For any $c_1,c_2>0$, $a<0$, and $b\in\mathbb{R}$ such that (s.t.) $a+b<0$, we have \cite[Eq. (06.35.21.0013.01)]{wolfrm}
\begin{align}\label{Lemma1}
\int_{c_1}^{c_2}{\rm e}^{bx}{\rm Ei}(ax)dx=\frac{1}{b}\left[{\rm e}^{bt}{\rm Ei}(at)-{\rm Ei}([a+b]t)\right]\!{\Big |}_{c_1}^{c_2},
\end{align}
where $f(t)|_{c_1}^{c_2}=f(c_2)-f(c_1)$ for the function $f(t)$.

\textbf{Proof:} Note that for $c_1,c_2>0$, $a<0$, and $a+b<0$, the exponential integral functions involved in both right- and left-hand sides (RHS and LHS) of \eqref{Lemma1} have negative arguments. In that case, based on the definition of ${\rm Ei}(x)$ for $x<0$ \cite[Eq. (8.211.1)]{gradshteyn2014table} we have $\frac{d}{dt}{\rm Ei}(lt)=-\frac{d}{dt}\int_{-lt}^{\infty}\frac{{\rm e}^{-x}}{x}dx={\rm e}^{lt}/t$, $\forall ~\!lt\!<\!0$. Then it is easy to verify that the derivative of ${\rm e}^{bt}{\rm Ei}(at)-{\rm Ei}([a+b]t)$ is equal to $b{\rm e}^{bt}{\rm Ei}(at)$; this completes the proof.

\textbf{Corollary 1:} $\forall c_1>0$, $a<0$, and $b\in\mathbb{R}$ s.t. $a+b<0$
\begin{align}\label{Cor1}
\int_{c_1}^{\infty}{\rm e}^{bx}{\rm Ei}(ax)dx=\frac{1}{b}\left[{\rm Ei}([a+b]c_1)-{\rm e}^{bc_1}{\rm Ei}(ac_1)\right].
\end{align}
\textbf{Proof:} Note first based on \cite[Eq. (2)]{harris1957tables} that ${\rm Ei}(-\infty)=\lim_{x\to-\infty} {\rm Ei}(x)=\lim_{x\to-\infty} {\rm e}^x/x=0$. Then by substituting $c_2=\infty$ in \eqref{Lemma1}, we have ${\rm Ei}([a+b]c_2)=0$ since $a+b<0$. Moreover, ${\rm e}^{bc_2}{\rm Ei}(ac_2)|_{c_2=\infty}=\lim_{x\to\infty} {\rm e}^{(a+b)x}/ax=0$. Therefore, \eqref{Cor1} can be inferred from \eqref{Lemma1} by inserting $c_2=\infty$.

\textbf{Corollary 2:} $\forall c_2>0$, $a<0$, and $b\in\mathbb{R}$ s.t. $a+b<0$ { \cite[Eq. (5.231.2)]{gradshteyn2014table} }
\begin{align}\label{Cor2}
\!\int_{0}^{c_2}\!\!\!{\rm e}^{bx}{\rm Ei}(ax)dx\!=\!\frac{1}{b}\!\left[{\rm e}^{bc_2}{\rm Ei}(ac_2)\!-\!{\rm Ei}([a+b]c_2)\!+\!\ln\!\left(\!1\!+\!\frac{b}{a}\!\right)\!\right]\!.\!
\end{align}
\textbf{Proof:} By substituting $c_1=0$ in \eqref{Lemma1}, we have ${\rm e}^{bc_1}{\rm Ei}(ac_1)-{\rm Ei}([a+b]c_1)=\lim_{x\to 0}{\rm Ei}(ax)-{\rm Ei}([a+b]x)$. Then using \cite[Eq. (1)]{harris1957tables}, $\lim_{x\to 0}{\rm Ei}(x)=\gamma+\ln|x|$, where $\gamma=0.57721$ is the Euler's constant. Therefore, for $c_1=0$ we have ${\rm e}^{bc_1}{\rm Ei}(ac_1)-{\rm Ei}([a+b]c_1)=-\ln\left(|a+b|/|a|\right)=-\ln\left(1+b/a\right)$. This completes the proof using \eqref{Lemma1}. Furthermore, note that by substituting $c_2=\infty$ in \eqref{Cor2} and using the discussions along with the proof of Corollary 1 one can obtain 
\begin{align}\label{Cor2_2}
\int_{0}^{\infty}{\rm e}^{bx}{\rm Ei}(ax)dx=\frac{1}{b}\ln\!\left(1+\frac{b}{a}\right),
\end{align}
which is the same result reported in \cite[Eq. (6.224.1)]{gradshteyn2014table}.

\section{Proof of \eqref{V1_avgI} and \eqref{Coeffs}}
In this appendix, we show how the expectation term involved in $V_1$ in \eqref{V_1} can be calculated in a closed form over $\boldsymbol{\mathcal{I}}_1$. To begin with, let us, for the ease of notation, define $\alpha_k\triangleq L'_kp'_k/(a_1L_1P)$, $k=1,2,...,K$, $B_{\tilde{g}}\triangleq [\sigma^2_{\mathcal{R}}+C_{\mathcal{D}}/\tilde{g}^2]/(a_1L_1P)$, $\overline{\boldsymbol{\mathcal{I}}}_{1,l}\triangleq\sum_{k=l+1}^{K}\alpha_k|\tilde{h}'_k|^2+B_{\tilde{g}}$, $l=1,2,...,K-1$, and $\overline{\boldsymbol{\mathcal{I}}}_{1,K}\triangleq B_{\tilde{g}}$. In order to take the expectation over $\boldsymbol{\mathcal{I}}_1$, we first average over $|\tilde{h}'_1|^2$ given that the other $|\tilde{h}'_k|^2$'s are some constant quantities, then average over $|\tilde{h}'_2|^2$, and so on. The expectation of the term involved in \eqref{V_1} over $|\tilde{h}'_1|^2$ can be calculated as
\begin{align}\label{V1,1}
&V_{1,1}\triangleq\nonumber\\
	&\!\E_{|\tilde{h}'_1|^2}\!\!\bigg[\!\exp\!\left(\!v_{1,i_1}\!\big[\alpha_1|\tilde{h}'_1|^2\!+\!\overline{\boldsymbol{\mathcal{I}}}_{1,1}\!\big]\!\right)\!{\rm Ei}\!\left(\!-v_{1,i_1}\!\big[\alpha_1|\tilde{h}'_1|^2\!+\!\overline{\boldsymbol{\mathcal{I}}}_{1,1}\!\big]\!\right)\!\!\bigg]\nonumber\\
&\stackrel{(a)}{=}\frac{1}{\alpha_1}{\rm e}^{\overline{\boldsymbol{\mathcal{I}}}_{1,1}/\alpha_1}\int_{\overline{\boldsymbol{\mathcal{I}}}_{1,1}}^{\infty}{\rm e}^{(v_{1,i_1}-1/\alpha_1)t_1}{\rm Ei}(-v_{1,i_1}t_1)dt_1\nonumber\\
&\stackrel{(b)}{=}\!\frac{1}{\alpha_1v_{1,i_1}\!\!\!-\!1}\!\!\left[\!{\rm e}^{\overline{\boldsymbol{\mathcal{I}}}_{1,1}/\alpha_1}{\rm Ei}\!\left(\!\!-\frac{\overline{\boldsymbol{\mathcal{I}}}_{1,1}}{\alpha_1}\!\!\right)\!\!-\!{\rm e}^{v_{1,i_1}\overline{\boldsymbol{\mathcal{I}}}_{1,1}}{\rm Ei}(\!-v_{1,i_1}\overline{\boldsymbol{\mathcal{I}}}_{1,1})\!\right]\!\!,
\end{align}
where step $(a)$ is derived by letting $f_{|\tilde{h}'_1|^2}(x_1)={\rm e}^{-x_1}$ and then $t_1=\alpha_1x_1+\overline{\boldsymbol{\mathcal{I}}}_{1,1}$, and step $(b)$ follows from \eqref{Cor1}. Note that in step $(a)$, $c_1=\overline{\boldsymbol{\mathcal{I}}}_{1,1}>0$, $a=-v_{1,i_1}<0$, and $a+b=-1/\alpha_1<0$, allowing to use Corollary 1 in \eqref{Cor1}. Using a similar procedure, the expectation of $V_{1,1}$ over $|\tilde{h}'_2|^2$ can be expressed as
\begin{align}\label{V1,2}
\!\!V_{1,2}\!\triangleq&\E_{|\tilde{h}'_2|^2}[V_{1,1}]\!=\!\frac{1}{\alpha_1v_{1,i_1}\!\!-\!1}\frac{1}{\alpha_2/\alpha_1\!-\!1}\!\bigg[{\rm e}^{\overline{\boldsymbol{\mathcal{I}}}_{1,2}/\alpha_2}{\rm Ei}\!\left(\!\!-\frac{\overline{\boldsymbol{\mathcal{I}}}_{1,2}}{\alpha_2}\!\right)\nonumber\\
&{\hspace{-0.4cm}}-{\rm e}^{\overline{\boldsymbol{\mathcal{I}}}_{1,2}/\alpha_1}{\rm Ei}\!\left(\!-\frac{\overline{\boldsymbol{\mathcal{I}}}_{1,2}}{\alpha_1}\!\right)\bigg]-\frac{1}{\alpha_1v_{1,i_1}\!-\!1}\frac{1}{\alpha_2v_{1,i_1}\!-\!1}\!\nonumber\\
&{\hspace{-0.4cm}}\times\!\bigg[{\rm e}^{\overline{\boldsymbol{\mathcal{I}}}_{1,2}/\alpha_2}{\rm Ei}\!\left(\!\!-\frac{\overline{\boldsymbol{\mathcal{I}}}_{1,2}}{\alpha_2}\!\right)\!-{\rm e}^{v_{1,i_1}\overline{\boldsymbol{\mathcal{I}}}_{1,2}}{\rm Ei}(-{v_{1,i_1}\overline{\boldsymbol{\mathcal{I}}}_{1,2}})\bigg].\!
\end{align}

For the ease of notation, let us further define ${\rm eEi}(x)\triangleq{\rm e}^x{\rm Ei}(-x)$, $\forall x>0$, and denote by $\beta^{(k)}_{\alpha_i}$, $i=1,2,...,k$, the coefficient of ${\rm eEi}(\overline{\boldsymbol{\mathcal{I}}}_{1,k}/\alpha_i)$ in $V_{1,k}$ after taking the $k$-th expectation, i.e., averaging over $|\tilde{h}'_k|^2$, $k=1,2,...,K$, and $\beta^{(k)}_{v_{i_1}}$ the coefficient of ${\rm eEi}(v_{1,i_1}\overline{\boldsymbol{\mathcal{I}}}_{1,k})$ in $V_{1,k}$. Using this notation, it is easy to observe that $V_{1,3}\triangleq\E_{|\tilde{h}'_3|^2}[V_{1,2}]$ can be expressed as
\begin{align}\label{V1,3}
\!\!\!V_{1,3}=&\beta^{(2)}_{\alpha_1}\frac{1}{\alpha_3/\alpha_1-1}\left[{\rm eEi}(\overline{\boldsymbol{\mathcal{I}}}_{1,3}/\alpha_3)-{\rm eEi}(\overline{\boldsymbol{\mathcal{I}}}_{1,3}/\alpha_1)\right]\nonumber\\
+&\beta^{(2)}_{\alpha_2}\frac{1}{\alpha_3/\alpha_2-1}\left[{\rm eEi}(\overline{\boldsymbol{\mathcal{I}}}_{1,3}/\alpha_3)-{\rm eEi}(\overline{\boldsymbol{\mathcal{I}}}_{1,3}/\alpha_2)\right]\nonumber\\
+&\beta^{(2)}_{v_{i_1}}\frac{1}{\alpha_3v_{1,i_1}\!\!-1}\left[{\rm eEi}(\overline{\boldsymbol{\mathcal{I}}}_{1,3}/\alpha_3)-{\rm eEi}(v_{1,i_1}\overline{\boldsymbol{\mathcal{I}}}_{1,3})\right]\!,\!
\end{align}
where, based on \eqref{V1,2}, $\beta^{(2)}_{\alpha_1}=-\frac{1}{\alpha_1v_{1,i_1}\!-\!1}\frac{1}{\alpha_2/\alpha_1\!-\!1}$, $\beta^{(2)}_{\alpha_2}=\frac{1}{\alpha_1v_{1,i_1}\!-\!1}\frac{1}{\alpha_2/\alpha_1\!-\!1}-\frac{1}{\alpha_1v_{1,i_1}\!-\!1}\frac{1}{\alpha_2v_{1,i_1}\!-\!1}$, and $\beta^{(2)}_{v_{i_1}}=\frac{1}{\alpha_1v_{1,i_1}\!-\!1}\frac{1}{\alpha_2v_{1,i_1}\!-\!1}$.

Using some inductive arguments and with the help of \eqref{V1,3} it can be shown that in the $k$-th expectation step, i.e., calculating $V_{1,k}$, we will have $k+1$ distinct terms of ${\rm eEi}(\overline{\boldsymbol{\mathcal{I}}}_{1,k}/\alpha_1),{\rm eEi}(\overline{\boldsymbol{\mathcal{I}}}_{1,k}/\alpha_2),...,{\rm eEi}(\overline{\boldsymbol{\mathcal{I}}}_{1,k}/\alpha_k)$, and ${\rm eEi}(v_{1,i_1}\overline{\boldsymbol{\mathcal{I}}}_{1,k})$. Therefore,
\begin{align}\label{V1,k}
V_{1,k}=\beta^{(k)}_{v_{i_1}}{\rm eEi}(v_{1,i_1}\overline{\boldsymbol{\mathcal{I}}}_{1,k})+\sum_{i=1}^{k}\beta^{(k)}_{\alpha_i}{\rm eEi}(\overline{\boldsymbol{\mathcal{I}}}_{1,k}/\alpha_i),
\end{align}
where the coefficients for $k=2,3,...,K$ can, recursively, be obtained as
\begin{align}\label{coefficients}
\beta^{(k)}_{\alpha_i}&=\frac{-\beta^{(k-1)}_{\alpha_i}}{\alpha_k/\alpha_i-1},~~~~~i=1,2,...,k-1,\nonumber\\
\beta^{(k)}_{\alpha_k}&=\frac{\beta^{(k-1)}_{v_{i_1}}}{\alpha_kv_{1,i_1}\!-1} +\sum_{i=1}^{k-1}\frac{\beta^{(k-1)}_{\alpha_i}}{\alpha_k/\alpha_i-1},\nonumber\\
\beta^{(k)}_{v_{i_1}}&=\frac{-\beta^{(k-1)}_{v_{i_1}}}{\alpha_kv_{1,i_1}\!-1}{=\frac{(-1)^k}{\prod_{i=1}^{k}(\alpha_iv_{1,i_1}\!-1)},}
\end{align}
with the initial values $\beta^{(1)}_{\alpha_1}=-\beta^{(1)}_{v_{i_1}}=(\alpha_1v_{1,i_1}\!-1)^{-1}$. 
{Finally, we need to prove \eqref{V1,k} with the coefficients in \eqref{coefficients} by recurrence. To this end, the base case is true by inserting $k=1$ in \eqref{V1,k} and \eqref{coefficients} and then recalling $V_{1,1}$ from \eqref{V1,1}. Moreover, assuming that the induction hypothesis (i.e., \eqref{V1,k} and \eqref{coefficients} for $k$) is true, we need to show that the induction step holds for $k+1$. Using the definition of $V_{1,{k+1}}\triangleq \E_{|\tilde{h}'_{k+1}|^2}[V_{1,k}]$ we have
\begin{align}\label{v1k+1}
V_{1,{k+1}}\!\!&\stackrel{(a)}{=}\!\E_{|\tilde{h}'_{k+1}|^2}\!\!\left[\beta^{(k)}_{v_{i_1}}{\rm eEi}(v_{1,i_1}\overline{\boldsymbol{\mathcal{I}}}_{1,k})+\!\sum_{i=1}^{k}\beta^{(k)}_{\alpha_i}{\rm eEi}(\overline{\boldsymbol{\mathcal{I}}}_{1,k}/\alpha_i)\!\right]\nonumber\\
&\hspace{-1cm}\stackrel{(b)}{=}\beta^{(k)}_{v_{i_1}}\E_{|\tilde{h}'_{k+1}|^2}\!\!\left[{\rm eEi}\!\left(v_{1,i_1}\left[\alpha_{k+1}|\tilde{h}'_{k+1}|^2+\overline{\boldsymbol{\mathcal{I}}}_{1,k+1}\right]\right)\!\right]\nonumber\\
&\hspace{-0.6cm}+\sum_{i=1}^{k}\beta^{(k)}_{\alpha_i}\E_{|\tilde{h}'_{k+1}|^2}\!\!\left[{\rm eEi}\!\left(\frac{1}{\alpha_i}\left[\alpha_{k+1}|\tilde{h}'_{k+1}|^2+\overline{\boldsymbol{\mathcal{I}}}_{1,k+1}\right]\right)\!\right]
\nonumber\\
&\hspace{-1cm}\stackrel{(c)}{=}\frac{\beta^{(k)}_{v_{i_1}}}{\alpha_{k+1}v_{1,i_1}-1}\!\left[{\rm eEi}\!\left(\overline{\boldsymbol{\mathcal{I}}}_{1,k+1}/\alpha_{k+1}\right)-{\rm eEi}\!\left(v_{1,i_1}\overline{\boldsymbol{\mathcal{I}}}_{1,k+1}\right)\!\right]\nonumber\\
&\hspace{-0.6cm}+\!\sum_{i=1}^{k}\!\frac{\beta^{(k)}_{\alpha_i}}{\alpha_{k+1}/\alpha_i-\!1}\!\left[{\rm eEi}\!\left(\frac{\overline{\boldsymbol{\mathcal{I}}}_{1,k+1}}{\alpha_{k+1}}\right)\!-\!{\rm eEi}\!\left(\frac{\overline{\boldsymbol{\mathcal{I}}}_{1,k+1}}{\alpha_i}\right)\!\right]\!,\!
\end{align}
where step $(a)$ follows from the induction hypothesis in \eqref{V1,k}, step $(b)$ is by recalling $\overline{\boldsymbol{\mathcal{I}}}_{1,k}\triangleq\alpha_{k+1}|\tilde{h}'_{k+1}|^2+\overline{\boldsymbol{\mathcal{I}}}_{1,k+1}$, and step $(c)$ is obtained using \eqref{Cor1} in a similar approach to \eqref{V1,1}. Hence, using step $(c)$ in \eqref{v1k+1}, we can characterize  $V_{1,{k+1}}$ as
\begin{align}\label{V1,k+1}
V_{1,{k+1}}&=\frac{-\beta^{(k)}_{v_{i_1}}}{\alpha_{k+1}v_{1,i_1}-1}{\rm eEi}(v_{1,i_1}\overline{\boldsymbol{\mathcal{I}}}_{1,{k+1}})\nonumber\\
&+\bigg[\!\frac{\beta^{(k)}_{v_{i_1}}}{\alpha_{k+1}v_{1,i_1}\!-\!1} \!+\!\sum_{i=1}^{k}\!\frac{\beta^{(k)}_{\alpha_i}}{\alpha_{k+1}/\alpha_i\!-\!1}\!\bigg]{\rm eEi}(\overline{\boldsymbol{\mathcal{I}}}_{1,{k+1}}/\alpha_{k+1})\nonumber\\
&-\sum_{i=1}^{k}\frac{\beta^{(k)}_{\alpha_i}}{\alpha_{k+1}/\alpha_i-1}{\rm eEi}(\overline{\boldsymbol{\mathcal{I}}}_{1,{k+1}}/\alpha_i),
\end{align}
which is equal to $\beta^{(k+1)}_{v_{i_1}}{\rm eEi}(v_{1,i_1}\overline{\boldsymbol{\mathcal{I}}}_{1,{k+1}})+\sum_{i=1}^{k+1}\beta^{(k+1)}_{\alpha_i}{\rm eEi}(\overline{\boldsymbol{\mathcal{I}}}_{1,{k+1}}/\alpha_i)$ with the coefficients defined similar to \eqref{coefficients} by substituting $k+1$ for $k$. This suggests that the induction step holds and completes the proof by recurrence. At the end, by substituting $k=K$ in \eqref{V1,k} and \eqref{coefficients} and noting that $\overline{\boldsymbol{\mathcal{I}}}_{1,K}=B_{\tilde{g}}$, one can obtain \eqref{V1_avgI} and \eqref{Coeffs}.}

\section*{Acknowledgment}
The authors would like to thank Mr. S. M. Azimi-Abarghouyi for the helpful discussions on an earlier draft version of this paper.

%


\end{document}